\documentclass[%
 reprint,
 amsmath,amssymb,
 aps,
]{revtex4-2}

\usepackage{graphicx}
\usepackage{dcolumn}
\usepackage{bm}
\usepackage{ulem}
\usepackage{xcolor} 

\begin{document}

\preprint{APS/123-QED}

\title{Emerging Physics Education Researchers' Growth in Professional Agency: Case Study}

\author{Shams El-Adawy}
\affiliation{Department of Physics, Kansas State University, Manhattan, Kansas, USA}

\author{Scott V. Franklin}
\affiliation{School of Physics and Astronomy, Rochester Institute of Technology, Rochester, New York, USA}

\author{Eleanor C. Sayre}
\affiliation{Department of Physics, Kansas State University, Manhattan, Kansas, USA}
\affiliation{Center for Advancing Scholarship to Transform Learning, Rochester Institute of Technology, Rochester, New York, USA}

\date{\today}

\begin{abstract}
Improving the physics enterprise to broaden participation in physics is one of the main goals of the physics education research community. Many classically trained physics faculty transition during their faculty career into engaging in research investigating the teaching and learning of their discipline. There is scarce research on the support and needs of these faculty as they engage in their first projects in this new research field for them. We investigate agency growth of two emerging physics education researchers and one emerging mathematics education researcher as they participate in a professional development program. We ground our case study analysis of interview data in a theoretical framework on agency. We identify the elements of the professional development program that were transformative in our case study participants ’ trajectory in education research. Receiving get-started information, building mechanisms to sustain research projects and engaging  with a supportive community help participants transform general  interests to specific questions, articulate concrete next-steps and increase their sense of self-efficacy. During this professional development program all three case study participants  gain agency in this new area of research for them. These identified program elements that affect agency growth can inform professional development opportunities for faculty transitioning into discipline-based education research, which expands our understanding of how to build capacity in the field.
\end{abstract}

\maketitle


\section{Introduction}
Physics education research (PER), a discipline-specific case of discipline-based education research (DBER), studies the human aspects of doing physics and aims to improve the teaching, learning, and inclusion of physics as a scientific field and human endeavor.  More generally, DBER ``investigates learning and teaching in a discipline using a range of methods with deep grounding in the disciplines' priorities, worldview, knowledge and practices"\cite{national2012discipline}. 

Because PER is physics, people who do PER are physicists \cite{aps992}, and thus studies of their learning, inclusion, and professional growth fall under the purview of PER.  How do people learn to do PER?  We are particularly interested in how classically-trained physicists in university faculty positions learn to engage in PER and develop their first research projects in this new-to-them subfield.  This is an important topic for understanding how our field grows and develops, alongside complementary studies of how graduate students become physics education researchers\cite{barthelemy2013they,van2014educational} and how our field has grown historically\cite{anderson2017linking}.


Studies of STEM faculty who engage in DBER have generally focused on Science Faculty with Education Specialities (SFES) \cite{bush2011investigation, bush2013widespread, bush2016fostering}. SFES refer to university science faculty who not only engage in research in science education, but also in other science education initiatives such as preparing future science teachers and course or curriculum development \cite{bush2017origins}. Most studies on SFES have focused on individual disciplines efforts in biology, chemistry, geoscience, mathematics, engineering and physics departments, rather than across STEM disciplines \cite{bush2020disciplinary, shipley2017transdisciplinary}. 

The literature on SFES faculty has reported on differences in the origin of SFES positions based on the type of institution the faculty are: at PhD-granting institutions, SFES are often hired to relieve other faculty from their teaching load; at MS-granting institutions, SFES are often hired to train future K-12 science teachers; and at primary undergraduate institutions (PUI), SFES often transition to their role after their hire to fulfill a need in their department \cite{bush2017origins}. The background and training of SFES faculty are varied and have been changing over the last decade, generally mimicking growth trends in PER over the last four decades \cite{anderson2017linking}.

The professional backgrounds of SFES can be divided into two rough groups.  One group, which is becoming more prevalent as graduate schools specifically prepare PhDs in DBER, have formal training in DBER via their PhD or postdoctoral training \cite{bush2019evolving}.  A second group were formally trained outside of DBER and transitioned to research on teaching and learning as part of their faculty appointment, either upon hire or years afterwards \cite{national2012discipline}. This second group of SFES are sometimes referred to as “boundary crossers” since they want to become more scholarly about teaching and learning, crossing the boundary between another subfield and education research.  Some maintain active research presence in both areas simultaneously, while others devote their entire scholarly effort to education research \cite{franklin2018peer}.  In either case, these SFES who begin to engage in education research as part of their faculty appointment are a group of emerging education researchers with limited formal opportunities to train in their new scholarly field.  

There is recent research studying challenges STEM education researchers face in finding community in DBER when those STEM faculty are trained in their specific discipline but have not been formally trained in DBER \cite{hass2021community}. There is also research that identifies the barriers these STEM faculty may face with other interdisciplinary education research such as the learning sciences community \cite{daniel2022challenges}. More research is needed on the type of support they need when entering the field and their growth as emerging STEM education researchers. In this work, it is important to center the connections among each emerging researcher's motivations in engaging in DBER, the constructs that are most relevant to each of them, and the experiences that sparked their interest in STEM education research. 

In this paper, we investigate emerging physics and math education faculty’s transition to DBER to examine the mechanisms by which we can best support them. We address the following research question:  \textbf{How do STEM faculty gain agency during the process of engaging in education research?} Exploring this question will allow us to better understand the ways in which DBER can be conducted in even more diverse instructional settings and institutions to improve STEM education. We explore this question through a multiple case study analysis of three participants in a professional development program, tracking how program activities affect their agency as researchers.

\section{Faculty agency in STEM}
We see faculty as agentic individuals who make changes to their teaching and adopt new teaching principles, building on their deep expertise in their institutional context as well as their pedagogical and physics knowledge.  This agentic lens has become a more common lens by which we investigate faculty professional development. This agentic perspective has been proven to provide valuable insight around their teaching because it highlights the strengths they bring to their teaching as both content and context experts \cite{el2022context, strubbe2020beyond}. 
The terms ``agency'' and ``professional agency'' have often been used interchangeably in DBER when investigating STEM faculty’s agency in instructional change. In particular, agency has often been examined from a particular lens: how individuals have agency within the power and constraints of systems they engage with and how they exert their agency \cite{gonzales2015faculty}. Less research has focused on faculty's perception of their agency.  In this section, we look at how agency has been studied in DBER, particularly in different STEM education fields and how we focus on perception of agency development in this study.  

In engineering education, professional agency growth and identity negotiation have been conceptualized in the context of instructional change to examine the enactment of professional agency of instructors given their individual resources and social conditions \cite{du2021engineering}. Du \textit{et al.} showed how instructors leveraged resources and social conditions to develop their agency and strategies to overcome pedagogical challenges they faced. 

In mathematics education, faculty’s professional agency has been studied in community colleges classrooms, where it was shown that part-time faculty are less agentive than full-time faculty in taking instructional decisions \cite{lande2016instructional}. Other studies focused on science and math instructors' professional development showed that teacher agency is a key feature in the successful implementation of professional learning communities \cite{ndunda2017university}.


In physics education, most research on agency and faculty has focused on different facets of  professional agency enactment or growth within the context of teaching. Strubbe \textit{et al.} use key parts of Bandura’s work to develop an analytical framework of faculty agency.  They characterize key features of physics faculty agency around their teaching to demonstrate the value of agentic-based perspectives in highlighting faculty's productive ideas they have about student learning and teaching \cite{strubbe2020beyond}. This research argues that researchers and educators should support faculty agency in their teaching, however, it does not provide the mechanisms by which we can facilitate that support.

Other research in physics education around agency in teaching include work on the Departmental Action Teams (DAT), the Faculty Online Learning Communities (FOLC), and the Physics and Astronomy New Faculty workshops (NFW), recently renamed the Faculty Teaching Institute (FTI) \cite{quan2019designing, dancy2019faculty, henderson2008promoting}.

DAT are externally facilitated groups consisting of faculty, students, and staff focusing on creating sustainable departmental change \cite{reinholz2019transforming}. The research team behind the DAT initiative highlights the value of having agency when they select an educational issue they will address \cite{corbo2016framework}. Nevertheless, DAT explicitly encourages hiring external facilitators for discussions on educational change, which can diminish faculty agency in leading change initiatives in their departments. 

FOLC are professional development communities built to support instructors in using research based teaching practices \cite{lau2021taxonomy}. Researchers engaging in studying FOLC have identified possible mechanisms to support physics faculty’s agency in teaching. This included focusing on their productive ideas to allow for deep reflection among faculty who participate in these conversations about teaching and educational change \cite{dancy2019faculty}. While FOLC are often lead by  facilitators, those facilitators are often drawn from participant near-peers, such as other faculty at similar institutions or former participants. 

NFW (renamed recently as Physics and Astronomy Faculty Teaching Institute (FTI) \cite{mckagan2020physport}) are workshops that aim to improve physics teaching by introducing new physics and astronomy faculty to research-based instructional strategies \cite{krane2010national}. In the research around the impact of NFW, researchers underlined the value of creating professional development experiences for faculty by helping develop their agency \cite{chasteen2020insights}. However, NFW has historically encouraged faculty to use the results and findings of physics education researchers in their classroom instead of engaging in doing research or curriculum development themselves, which directly depresses physics faculty’s agency in instructional change. 

Each of these programs and initiatives address particular physics faculty needs when it comes to instructional and departmental changes: multi-layer facilitation of department instructional change (DAT), community building and peer support in implementing research-based practices (FOLC) and advocacy in using research-based practices (NFW). The research on these program highlights the value of agentic perspectives when examining STEM faculty instructional change.  These programs address the faculty as instructors or as department members who perform service, but they do not address -- and in some cases, specifically exclude -- faculty acting as researchers.  

Less research has focused on faculty interested in doing scholarly work on learning and teaching rather than solely implementation of research results.  Hence, we fill a gap in the literature by investigating the perception of agency growth of physics and math researchers transitioning into education research. For STEM researchers who have not received extensive training in discipline-based education research, we have minimal evidence of their experiences as they transition into DBER. In particular, our context draws participants from multiple institutions (similarly to NFW) and promotes community development among participants with shared goals (similar to FOLCs).  Our case study participants engage in program activities with the intent to bring their growing skills and research projects back to their home institutions. 

\section{Theoretical Framework: Agency}
The literature on student and faculty agency is extensive. One of the most commonly used frameworks for studying agency of both student and faculty comes from social cognitive theory. From this lens, human agency theory of development, adaptation and change  adopts the view that individuals are products of the interplay among interactions of environment, behavior and self \cite{bandura_functional_2012}. Bandura’s agency lens informs part of the subject-oriented socio-cultural approach to agency growth which posits that the subject of focus is an agentic actor in relation to the social world  and agency is temporally constructed within engagements with different tasks \cite{etelapelto2013agency}.

Bandura’s work lays some of the foundation for how the definition of professional agency came to be. According to Etelapelto, “professional agency is practiced when professional subjects and/or communities exert influence, make choices and take stances in ways that affect their work and/or their professional identities” \cite{etelapelto2013agency}. The concept of professional agency focuses on agency in one’s career and has been studied in the context of instructional change and teacher education.  The four main areas are professional development, education policies, teacher identity development and social justice \cite{hinostroza2020university}. These  topics overlap and interact with each other as they constitute the major factors that provide affordances or constraints in university faculty’s professional agency growth. 

In this paper, we use Bandura's agency framework with a focus on professional agency examining how participants perceive their agency development during their engagement in a professional development program as they transition into DBER.  In other words, within the context of professional development, we use Bandura's definition of agency as an individual’s ability to make choices and take action based on intentionality, forethought, self-reactiveness and self-reflectiveness \cite{bandura_social_2001} to ground our analysis of STEM faculty who transition to DBER. Table \ref{tab:definitions} summarizes the definition of the components from Bandura’s framework and how they are operationalized within the context of this study. 

\begin{table*}[hptb]
    \centering
    \begin{tabular}{p{0.2\linewidth}p{0.4\linewidth}p{0.4\linewidth}}
        \textbf{Agency components} & \textbf{Theoretical definition} & \textbf{Contextual definition} \\ 
        \hline

        Intentionality & Planning for specific actions for the short or long terms to achieve goals & What emerging STEM education researchers plan to do or accomplish with their first DBER research projects \\ 
        \hline

        Forethought & Process of setting goals,  anticipating actions and  consequences to reach desired outcomes  & What research tasks emerging STEM  education researchers are considering undertaking  and what they anticipate they need to successfully  complete their first DBER research project \\

        \hline
        Self-reactiveness & Motivation and self-regulation needed to execute actions planned  & What interest in DBER emerging STEM education researchers discuss, especially what drives their intrinsic motivation to engage in DBER\\ 
        \hline

        Self-reflectiveness & Belief in one's perceived competence in their ability to undertake a behavior (self-efficacy) & What emerging STEM education  researchers perceived competence in DBER to be \\ \hline 
      \end{tabular}
    \caption{Definition of the components from Bandura’s framework and how they are contextextualized in this study}
    \label{tab:definitions}
\end{table*}

Intentionality refers to planning by setting measurable steps for the short or long term to achieve specific goals.  When engaging in any research project, deliberate thinking and mapping of the research directions is a common practice. Research has shown that intentionality plays a critical role in mentorship, especially in experiential learning experiences \cite{thomas2008preparing}. So when engaging in a first DBER research project, articulating intent and the scale of engagement can help solidify research directions. In our context, we are using it to get a deeper understanding of what emerging STEM education researchers are aiming for by engaging in DBER and how their aims and hopes evolve as they engage in an experiential learning experience. 

Forethought, the process of defining specific tasks and their potential impact on target goals, is an integral part of the research process. Forethought allows us to articulate deeper details that provide insight into intentionality. Researchers show how critical the task analysis/strategic planning phase of forethought plays in self-regulating behaviors and motivation in the short and long term \cite{english2013supporting}. In our context, DBER tasks planning allows us to get a deeper understanding of how emerging STEM education researchers plan to engage in DBER. It provides more details than intentionality and connects with the other components (self-reactiveness and self-reflectiveness). By exploring emerging STEM education researchers' forethought, we can identify what tasks and behaviors they envision needing most to support them in their DBER research project.

Self-reactiveness is the motivation and self-regulation needed to execute a planned action. In our context, we are particularly interested in emerging STEM education reserachers’ intrinsic motivation to DBER. We are drawing on Self-Determination Theory (SDT), a theory about motivation that centers around a learner’s agency when making choices to reach desired goals \cite{ryan2000self} to examine, justify and interpret their development.  SDT suggests that three psychological needs: competence, relatedness and autonomy have to be satisfied to have the most self-determined form of motivation \cite{stupnisky2018faculty}. In our context competence refers to the need to feel proficient in engaging in DBER. Relatedness refers to the need to feel connected to the DBER community, the people and the research products value.  Autonomy refers to the need to have a sense of choice in behavior and tasks that drive their engagement in DBER. Although we are most interested in understanding intrinsic motivation, SDT draws us to also examine the role of extrinsic motivation, especially how external factors can regulate behavior. These emerging STEM education researchers are engaging in DBER while holding other responsibilities within their respective institutions, which inevitably plays a role in how they view their role in DBER. As such, understanding DBER engagement through the interplay of these regulatory factors can provide insight into their motivation to do DBER and how it evolves as they engage with a DBER professional development program. 

Self-reflectiveness refers to a “self reflective belief in one's ability to succeed”. In our context, DBER self-reflectiveness allows us to gain insight into emerging STEM education perceived abilities to do research, which can help us understand what support is needed. When looking at self-reflectiveness, Bandura draws our attention to self-efficacy, which is defined as one’s perceived skill and competence in their ability to undertake a behavior \cite{bandura_self-efficacy_1982}.  Understanding emerging DBER self-efficacy can inform support of what professional development needs would be most beneficial. Bandura’s work investigating the relation between agency and self-efficacy underlines how changes in self-efficacy have a direct and critical impact on agency, where increasing self-efficacy is a necessary condition for increasing agency, whereas increasing other aspects of agency does not necessarily entail an increase in self-efficacy \cite{bandura_functional_2012}. Bandura’s theory on self-efficacy also suggests four different sources that contribute to a person’s perceived self-efficacy: mastery experiences, vicarious learning, verbal persuasion, and physiological states \cite{bandura_self-efficacy_1982}. In our context, mastery experiences refer to experiences that provide information about personal successes or failures in task similar to the new DBER experience they are engaging in and influence emerging STEM education researchers’ confidence in their ability to perform a DBER related task.  Vicarious learning refers to learning that occurs by observing others performing DBER either by observing how they are engaging in DBER or how they compare their DBER work with others. Verbal persuasion is related to messages received about their ability to do DBER conveyed through interactions with the DBER community. Physiological state refers to emotional indicators that an emerging STEM education researcher may rely on when evaluating their ability to do DBER.  

These four elements of agency: intentionality, forethought, self-reactiveness and self-reflectiveness, are  interrelated and combined they provide us with a valuable and holistic lens to study the factors that impact STEM faculty as they transition into DBER. 

\section{Data}
The backdrop for this research study is a professional development program, Professional development for Emerging Education Researchers (PEER), designed to help faculty, postdocs and graduate students jumpstart their transition into the world of discipline-based education research \cite{franklin2018peer}. The central activity of PEER is a series of workshops to help participants design and conduct research projects; engage in targeted experiential work to develop their projects and skills; and collaborate and form a support community of peers, mentors and collaborators.

Conducting this study within the context of PEER was advantageous for two main reasons: firstly, our data collection was grounded in participants' experiences with the program; secondly, most participants were leading their first research projects in DBER so it was an opportune time to examine the mechanisms that best support emerging STEM education researchers.

For this particular study, the data stems from participants in one of the virtual editions of the PEER program, primarily drawing from a national audience of emergent mathematics and physics education researchers. This field school occurred over Zoom through spring 2021 and was attended by 45 emerging mathematics and physics education researchers from a variety of research and teaching institutions across the US. The workshop consisted of a kickoff session, three two-hour sessions spread over 6 weeks, and then a three-day intensive at the end of June. Table \ref{table2:peeractivities} lists goals and activities of each set of workshop sessions during PEER. As can be seen in the table, PEER is professional development program expecting active engagement from participants where collaboration, responsivity and group work is embedded in all aspects of the program. At the end of each session, participants were asked to list their questions they had and topics they wanted to learn more about. The following sessions incorporated these questions and interest participants had.

\begin{table*}[ht!]
\caption{Goals and activities of each set of workshop sessions at PEER}
\begin{tabular} {p{2.8cm} p{7cm} p{7cm}} \hline
{} &{\bf Goals} & {\bf Activities } \\ 
\hline
Kickoff Session &  Introduce facilitators and participants to establish community and to set expectations and norms about the series of workshops. & Topical discussions: ethics of collaboration

Generative writing \\ \hline Session 1 & Work on research design by identifying key features of a research question and recognizing that they change over time and refining it through collaborative feedback and self-reflection on their particular contexts for research and data access. & Topical discussion and procedural knowledge: research questions and data access 

Generative writing \\ \hline Session 2 & Articulate differences between research models and implications for practice as well as  develop a plan for the next steps of your research project. & Topical discussions and procedural knowledge: models of research and topic of interest for participants: literature reviews, data collection and IRB 

Generative writing \\ \hline Session 3 & Provide feedback on individual projects and help participants understand the impact of their worldview on their research process and research project & Individualized project feedback from facilitators 

Topical discussions on theoretical frameworks 

Generative writing  \\ \hline 3-Day Intensive & Engage in observational data analysis, theory discussion, individual project feedback and dissemination of work and planning for next steps post-PEER. & Individualized project feedback from facilitators

Topical discussions and procedural knowledge:  observational data, theory and dissemination of research

Project mapping of goals with specific tasks for both the near (days, weeks) and far (months) future

Generative writing \\ \hline
  \end{tabular}
  \label{table2:peeractivities}
\end{table*}

Participants were solicited for semi-structured interviews before and after participation. Semi-structured interviews are a common tool for data collection in qualitative research, which uses a series of open-ended questions allowing for emerging themes in the discussion to be explored \cite{harvey2001process}. Thirteen participants were interviewed pre-participation and eleven participants were interviewed  in the post-interviews. Only five participants took part in both pre and post interviews (for a total of nineteen participants). 

\section{Methods}
A case study as a methodological approach focuses on developing an in-depth analysis of a case or cases to capture the complexity of the unit of analysis \cite{yin2009case, creswell2016qualitative}.  When doing a multiple case study, the researcher selects a few case study participants to illustrate the unit of analysis, which is agency.

In practice, the case study analysis was conducted as follows. The first author conducted a preliminary analysis of the interview data, which highlighted participants who addressed the unit of analysis. Three of the five participants who took part in both pre and post interviews were chosen using a purposeful sampling process,  a common way of selecting cases in qualitative research \cite{suri2011purposeful, emmel2013purposeful}.  Our case study participants  addressed components of agency and were at different stages of their faculty career as they started to engage in their first education research project in their discipline. Given that these faculty were at a similar stage of engagement with DBER but at different stages of their professional lives, they allow us to investigate variation and similarities in agency growth in education research for faculty with different teaching and research experiences. 

After selection, the first author provided a detailed description of themes for each case study participant grounded in Bandura’s agency framework. This thematic analysis was expanded to compare and contrast themes across cases to identify the key elements relating to agency growth. After the themes across cases were characterized in the theoretical framework, the first author wrote an initial analysis of the case studies. Then, members of the research team reviewed the analysis together. The instances where disagreement was identified, a discussion followed until agreement was met. This  process often resulted in the first author reviewing the interview transcripts to provide more evidence for their interpretation with the relevant pieces of data. 

\section{Case Study Participants}
Our three case study participants  are given the pseudonyms Olivia, Madison and Akemi. During their participation in this study, Madison and Akemi are leading their first physics education research (PER) project, whereas Olivia is leading her first math education research (MER) project. As they are chairing the first project in discipline-based education research, all three case study participants identified a common need to gain practical skills and a better sense of what the field of DBER was. These beliefs and motivations were a strong reason for their participation in a professional development program such as PEER. 

Olivia is a Full Professor in a mathematics department at a public land grant university. She has been in her current mathematics departments for over twenty years teaching introductory and upper-level mathematics courses. Her graduate training is in mathematics and her current primary research area is in graph theory. During her participation in this study, she was exploring mathematics education research to help make evidence based instructional changes in her classroom and institution. In the course of her engagement with PEER, Olivia focused on developing her existing research project and developed an understanding of where she can situate herself in mathematics education research (MER). 

Madison is an Associate Professor in a physics department at a primarily undergraduate institution. She teaches many of the undergraduate physics courses, but she is especially passionate about instructional laboratory teaching in physics. Her graduate training is in physics and her current primary research area is in condensed matter physics. During her participation in this study, she started exploring physics education research to inform and assess her work redesigning instructional labs in her department, which included facilitating a departmental faculty learning community. In the course of her engagement with PEER, Madison focused on narrowing down her research questions and getting started on writing an NSF grant to fund her physics education research (PER) project. 

Akemi is a Visiting Faculty Member in a physics department at a private liberal arts college during the pre-interviews and a high school science teacher in the post-interview. She was teaching a few introductory undergraduate physics courses and was about to teach high school science. Her graduate training is in physics with a research focus in condensed matter physics. As an early-career scientist during her participation in this study, she was engaging in physics education research to make evidence-based instructional decisions in the classroom as well as build her research portfolio for her career advancement. In the course of her engagement with PEER, Akemi focused on refining her research project and getting started with data collection. 
\section{Analysis}
In the following section, we discuss our data analysis within each component of Bandura's framework for each case study participant. We present each participant's pre-PEER status and post PEER status for each component of the framework. 
\subsection{Intentionality}
Intentionality in the literature is defined as the planning for specific actions for the short or long terms to achieve goals. In our context, intentionality refers to what emerging STEM education researchers plan to do or accomplish with  their first DBER research projects. Features of intentionality were brought up by participants pre and post-PEER, highlighting the alignment of short and long-term plans with motivation. Project mapping at PEER was the central common activity contributing to  intentionality. Table \ref{tab:intentionality} summarizes the status of intentionality for each participant. 

\begin{table*}[tbh]
\caption{Status of intentionality (plans) pre and post PEER for each participant}
  \begin{tabular}{p{3cm} p{7cm} p {7cm}}
\hline
{\bf Participants} &{\bf Pre-PEER} & {\bf Post-PEER} \\ 
\hline
Olivia & {\bf{Alignment of short and long-term plans with motivation}}: 
\linebreak
Short-term: complete her MER project about assessing the impact of a new pedagogical strategy
\linebreak Long-term: keep finding ways to measure how to make learning math better for students in math courses at her institution
 & {\bf{Refined conceptualization of short-term plan:}}
 
Short-term: submitting a paper and attending an upcoming MER conference \linebreak
Long-term: keep finding ways to measure how to make learning math better for students in courses at her institution. \\ \hline
Madison & {\bf{Alignment of short and long-term plans with motivation:}} \linebreak
Short-term: go through and complete at least one iteration of the research design of a PER project
\linebreak Long-term: PER fits into her long-term career trajectory where she hopes to incorporate PER in her research portfolio & {\bf{Refined conceptualization of short-term plan:}}

Short-term: obtain a National Science Foundation (NSF) grant for her PER project
\linebreak Long-term: PER fits into her long-term career trajectory where she hopes to incorporate PER in her research portfolio
 \\ \hline
Akemi & {\bf{Alignment of short and long-term plans with motivation:} }

Short-term: complete her current PER  project to have a DBER project to discuss when applying for more permanent jobs
\linebreak Long-term: PER fits into her career prospects
 & {\bf{Refined conceptualization of short-term plan:}} \linebreak
Short-term: submission of conference paper and a longer journal paper for her current PER project
Long-term: PER fits into her career prospects

\\ \hline
  \end{tabular}
  \label{tab:intentionality}
\end{table*}

\subsubsection{Pre-PEER Intentionality }
Olivia’s short term plan is to complete her MER project about a new pedagogical strategy they are implementing in their calculus courses, which is specific to her institutional context and issues they are having about introductory math courses:
\begin{quote}
    \textit{We are basically open admissions, which means we do get a lot of students who are first generation, low income, and so have nonmathematical readiness issues, [...] so looking at whether or not taking those students with really low prerequisite skills and putting them in a class that’s going to provide them with this corequisite support over the course of calculus, that is, to not ask them to drop back to precalculus but keep them in calculus with a little extra support. And we’re going to measure whether or not these students with low scores look like they can be successful in calculus.}
\end{quote}
By providing background information about the type of institution she is at, Olivia sets the stage for her intention she articulates, which is measuring the impact of providing additional support to students in calculus instead of dropping them to pre-calculus courses. 
Her long term plan is to keep finding ways to measure how to make things better for students in math courses at her institution by investigating different instructional change strategies. Her intentions to engage in MER is to improve passing and retention rates in math courses. She is supported by her institutional context where she hopes to implement effective instructional strategies. 

Comparable to Olivia, Madison’s short term plan is to go through and complete at least one iteration of the research design of a PER project in order to be able to write and submit a grant proposal:
\begin{quote}
    \textit{I would just really love to come out with some type of completed product of, even if it’s like a draft or a logic module or, you know, questionnaires that I can send out. Start getting some concrete documentation and things to prepare for grant submission for this.}
\end{quote}
PER also fits into Madison’s long-term career trajectory, where as a tenured professor, she hopes to incorporate PER in her research portfolio:
\begin{quote}
    \textit{Basically, this is my first year as a tenured associate professor, so a question is always what do I do now? What is going to be my next big thing to get me from associate to full professor? And I have my materials research, and I’ll continue to do that, but I really like the idea of kind of adding on to what I do as a way to get myself to that next big step in my career.}
\end{quote}

Likewise we have similar factors as part of Akemi’s intentions in doing DBER. On the short-term scale, Akemi hopes to complete her current PER project to have a DBER project to discuss when applying for more permanent jobs:
\begin{quote}
    \textit{I have a major goal is kind of find what improves or decreases my students’ self-efficacy [...]I’m a visiting professor, so I’m not required to do research, but I know I have to do research to get a better job.}
\end{quote}
On the long term scale, PER fits into her career prospects as she is looking for jobs that will require her to do research. She is considering faculty positions and public engagement positions where doing research on instructional change and best practices would be a core component. Doing and completing a STEM education  research project will provide evidence of her expertise when she applies to more permanent positions, which will aid her career advancement. 

Pre-PEER in intentionality, we identified that on the long term scale Madison and Akemi wanted PER to become a major component of their research portfolio. Whereas Olivia wanted to keep being involved in improving teaching at her institution and sees that engaging in MER will allow her to do so. To reach those long-term goals, all three case study participants wanted to complete at least an iteration of the research design process.

\subsubsection{Post-PEER Intentionality}
The process of project mapping helped Olivia identify specific goals and actions that needed to occur and set intentions for the short term, which include submitting a paper and attending an upcoming MER conference:
\begin{quote}
    \textit{It both helped me make specific plans to submit a paper, clarify a new research question that is both new to me but also a new kind of skill set I need to address it [...] it helped me clarify what it is, is sort of my more recent research project, and it’s also helped me produce a more specific set of future plans. So, you know, submitting a particular paper, attending a particular conference, that sort of thing.}
\end{quote}
In the long term, she plans to continue asking similar questions about improving math education. She plans to remain intellectually engaged in MER by creating master’s students research projects with data analysis relevant to her research:
\begin{quote}
    \textit{I’m probably kind of committed to a series of short-term plans for now. The other thing that I’m also in a really advantageous position is that my department has a statistics master’s program [...] If I can produce data from our institutional database, I can implicitly produce a statistics project for a master’s student who needs a statistics project, and so that also helps me kind of keep thinking about some of my questions.}
\end{quote}
We see in the long-term plans, there is no significant difference in her intentions, but we can notice that project mapping helped refine conceptualization of short term plans. 
For Madison, her short and long term plans have not changed. However, she articulates more concrete steps in achieving those goals, which came up when she discussed the planning for next steps that happened at PEER. She wants to apply for a PER grant:
\begin{quote}
\textit{I’m hoping over the next six months to work on a grant and that’s going to be various steps. I want to start actually just making like visuals for it just to help me process like what is the flow of the project, like the logic models and stuff. And yeah, the big goal for me is one of the like NSF education grants.}
\end{quote}
Moreover, she still wants to include PER as a main area of research portfolio and is considering dedicating her sabbatical to this endeavor:
\begin{quote}
    \textit{Oh, the other like longer, longer term planning thing as well is part of the grant is my kind of strategic career plan. I got a sabbatical I could take at some point, so I would really like to have the money to take like a full year sabbatical to really focus on the education research side.}
\end{quote}

For Akemi, her short term plan is more specific compared to the pre-interviews. Taking the time to map out her research projects at PEER enabled her to realize that she is at a stage where she is trying to find a good journal home. She is considering submission for a short conference paper and a longer journal paper: 
\begin{quote}
    \textit{I can try a two-page proposal to the International Conference of Learning Science, I believe. That’s-[...]So try that and then see how it goes, and then after that, I can try something like PRPER [Physical Review-Physics Education Research], writing a 15-to-20-page stuff. So that’s my goal, so I’m trying to put that two-page thing, and then see what I missed.}
\end{quote}
Her long term plan is still related to her job prospects but she switched positions during PEER. She is now a high school science teacher and may consider keeping her current position, but it is unclear where PER will fit into that:
\begin{quote}
    \textit{So I think the thing I’m imagining is more like I collaborate with someone else, and they probably teach at university or college and then I might… Then, I don’t know, if I’m researching on my own students, I don’t think the IRB will review that. I’m interested in that, but I don’t know a way to research on high school students.}
\end{quote}

Post-PEER intentions in the short and long terms to improve teaching practices by doing DBER remained the same for Olivia, Madison and Akemi. Nevertheless, they had a more defined trajectory on how they will keep engaged in DBER work post-PEER, especially in their short term planning. The most significant PEER activity in their refinement of short term plans was the project mapping that happened at PEER that allowed each participant to conceptualize the next steps of their projects. 

\subsection{Forethought}
In the literature, forethought is defined as the process of setting goals, anticipating actions and consequences to reach desired outcomes. In our context, forethought refers to what research tasks emerging STEM education researchers are considering undertaking and what they anticipate they need to successfully complete their first DBER research project.  Unlike intentionality where the participants brought up project mapping, several program activities were brought up by participants when it came to forethought: interactions with facilitators, topical discussions, DBER literature, procedural knowledge and project feedback. However, the common theme pre-PEER  in forethought was the common need for research project design support. Given the different stages they were at pre-PEER, there were nuances specific to each case participant's DBER project in the actions they foresee and post-PEER these were refined with the nuances relevant to each.  Table \ref{tab:forethought} summarizes the status of forethought pre and post PEER for participants.

\begin{table*}[tbh]
\caption{Status of forethought (research tasks)  pre and post PEER for each participant}
  \begin{tabular}{p{3cm} p{7cm} p {7cm}}
\hline
{\bf{ Participants} }& {\bf Pre-PEER }  &{\bf Post-PEER} \\ 
\hline
Olivia & 

{\bf{Articulates need for nuanced research project design support:}}

Articulates need for research question refinement

 &
{\bf{Articulates refinement of research project design with nuances:}}

Articulates where her research question fits within the MER field and how to move forward

Anticipates potential challenges related to data types and analysis she uses

\\ \hline
Madison  & 
{\bf{Articulates need for nuanced research project design support:}}

Articulates need for research question refinement and structure to move research project forward

&

{\bf{Articulates refinement of research project design with nuances:}}

Articulates how she transformed her general PER interest into a viable PER research question and project

Articulates specific research tasks she is engaging in: grant writing

Articulates need to continuously engage in professional development that helps her move her research project forward

\\ \hline
Akemi & 
{\bf{Articulates need for nuanced research project design support:}}

Articulates need for research question refinement, need for structure to move project forward and need for guidance on data analysis

& 

{\bf{Articulates refinement of research project design with nuances:}}

Articulates how she transformed her general PER interest into a viable PER research question and project

Articulates specific research tasks she is engaging in: data collection and analysis

\\ \hline
  \end{tabular}
  \label{tab:forethought}
\end{table*}

\subsubsection{Pre-PEER Forethought}
Olivia has little interactions with the DBER community, even informally, however she had submitted a proposal to the National Science Foundation (NSF) to examine the ways in which those with very little perquisite skills succeed in calculus class with additional support. She had identified many parts of her research process and identified the areas she believed she needed most help, which are refinement of research questions  and writing of a science education grant. She is anticipating refining her research questions by engaging with researchers with various backgrounds in the field.
She is also anticipating the need to get a broad view of MER and the different steps of the research design process to enhance her grant writing: 
\begin{quote}
    \textit{So a lot of that sort of the nuts-and-bolts aspects of submitting a science education, math education or sort of community transformation sort of grant is clearly, I’m clueless and I could use help on that.}
\end{quote}

In parallel, Madison has had several informal conversations with members of the DBER community. She collaborated with education researchers when she opened up her classroom for data collection for education projects. She has concrete ideas for her research project and knows she needs help refining her research question and project into a tangible and viable study. She anticipates the need for guidance with different steps of the research design process. In particular, articulating and refining her research interest into a viable PER project:
\begin{quote}
    \textit{I’m coming in with kind of a concrete idea, I would love it if it’s almost like stepping me through what the project should look like. Like, helping me take what I have and think of ways of okay, how do I go with this kind of nebulous idea of faculty learning communities and labs, and how do I do take all these steps we’re going to talk about, like how do you assess the grant, how do you come up with good research.}
\end{quote}

Unlike Madison, Akemi has had few interactions with the DBER community, but she is collecting data in her classroom to pursue her research interest. She has identified the need to better understand the structure of how to conduct DBER research as someone unfamiliar with the research field. She put together a proposal to conduct a research study to promote equity in her physics classroom but wanted guidance on refining her research questions. She also highlights needing help with data analysis to move forward with her research:
\begin{quote}
    \textit{ I do want to learn how to analyze my data, I believe there is something about like coding stuff like that, but I don’t really know how to code my data. So yes, that’s definitely something that I want to learn.}
\end{quote}

In forethought pre-PEER, we examined the common need that all three participants identified: research design refinement, particularly the refinement of their research questions. However, there were nuances in their stages of the research process. Olivia applied for a grant from the NSF so she had made an attempt to identify all the different parts of the research design process. Akemi had put together a PER study proposal at her institution that was accepted and she was already in the data collection phase. Madison had interacted peripherally with PER projects by welcoming researchers into her classroom  to collect data. She had identified a research interest, but she had not put together the pieces of her research design. 

\subsubsection{Post-PEER Forethought}
For Olivia, by interacting with other STEM researchers through the workshops and the MER literature, she describes having a better sense of the DBER community. Readings and interactions with other participants at PEER has broadened her understanding of MER as a mathematician.  It has also broadened her perception of what is MER, who does MER and how she perceives the field. She foresees herself playing a useful role bridging the disconnect that can exist between mathematicians and mathematics education researchers. 
PEER has also helped her articulate some specific actions and consequences she is anticipating as she continues to move forward with her MER project. Although she anticipates finding time to do DBER in her schedule challenging, she views time constraints as keeping her accountable. She will be encouraged to continue her interactions with the DBER community in the near future to complete her current MER project:
\begin{quote}
    \textit{I’m going to be forced for the next three years to be reaching out to DBER people in some form. And sort of periodically reevaluating whether or not I’m reaching my goals, not just with the project but more broader, like the things I specifically talked about, keeping in contact with people I’ve met and continuing to broaden my reading.}
\end{quote}
She also anticipates some criticism of her work from the broader DBER community because her quantitative analysis does not depict a complete picture of students’ progress in their math courses and she anticipates the need for qualitative lens. For Olivia,  the PEER activities that played a role in her growth in forethought were interactions with facilitators with DBER expertise, topical small group discussions and guiding engagement with DBER literature.

As for Madison, different elements contribute to her growth in forethought. She has a better sense now of what a viable education research process is and how to transform research interest into a research project in PER. She foresees seeking out similar interactive professional development  programs that focus on participants’ specifics research projects:
\begin{quote}
    \textit{[PEER  helped identify] how do you go about transforming something you might be curious about into something that’s a viable research project? And I really like that. [...]Even if it was just like webinars or something, I would love to continue to engage with this because I feel like the workshops were… I like that they were really interactive, I like that they gave us a lot of time to work on our projects ourselves and I would love to do that in like a more guided sense.}
\end{quote}
Obtaining get-started information has provided her with a roadmap on how to move from research interest to a viable research question and project in DBER. She has learned how to articulate her research question and refine it through the many successive opportunities in the PEER workshops that broke down the tasks related to DBER into manageable pieces. In particular,  the PEER activities that played a role in her growth in forethought were procedural knowledge workshops, topical small group discussions and individualized project feedback from facilitators. 

For Akemi, obtaining information on how to get started has provided her with a roadmap on how to move her research project forward. She has gained insight into the significance of the different parts of the research design process such as theory and limitations. She values how explicitly DBER people think about the limits of their understanding, which she did not see much of in condensed matter physics.  She believes that she has a far better idea of what her project is and how to move forward with it, as compared to before doing PEER. To situate her work within the field, she anticipates framing her papers in a similar structure to the PER literature she has been engaging with. She anticipates being better at assessing work related to her research topic because she has a good solid background on the foundational work in her research interest area:  
\begin{quote}
    \textit{I think, I mean, if I see some new theory, I’ll definitely pay more attention about to learn that. If they are talking about self-efficacy and they are not using Bandura, it’s like Bandura is everywhere and then so currently, I haven’t found anything new on the theory that I’m doing, but if I find a different one, I would use that as a keyword to find more paper.}
\end{quote}
 For Akemi,  the PEER activities that played a role in her growth in forethought were mainly  procedural knowledge workshops and gudiment engagement with DBER literature. 

Workshop structure, content and community at PEER helped research project refinement for Olivia, Madison and Akemi. For Olivia, workshops that addressed refinement of research questions iteratively, setting specific DBER plans for the near future addressed aspects of the research design she had identified needing pre-PEER.  For Madison, the iterative process of the research design that participants went through at PEER helped her refine her research design. She has also identified more specific parts of the research design she will want help with in the future. For Akemi, through readings and discussions, she learned about the norms of the field and how to situate and shape her project within it.

\subsection{Self-reactiveness}
In the literature, self-reactiveness refers to motivation and self-regulation needed to execute actions planned. In our context, self-reactiveness refers to what interest in DBER emerging STEM education researchers discuss, especially what drives their intrinsic motivation to engage in DBER.
Similarly to forethought, a common interest motivates participants in engaging in DBER and multiple program elements affect growth in self-reactiveness. However, more program elements are highlighted and impact the nuances of self-reactiveness post-PEER than forethought.  Table \ref{tab:selfreactiveness} summarizes the status of  self-reactiveness pre and post PEER for participants.

\begin{table*}[tbh]
\caption{Status of self-reactiveness (sources of motivation) pre and post PEER for each participant}
  \begin{tabular}{p{3cm} p{7cm} p {7cm}}
\hline
{\bf{ Participants} }& {\bf Pre-PEER } &{\bf Post-PEER} \\ 
\hline
Olivia & 

{\bf{Competence and relatedness}} drive motivation to improve teaching by doing DBER.

 &
{\bf{Competence and autonomy}} still drive motivation to improve teaching by doing DBER; refined conceptualization of short-term plan.

\\ \hline
Madison  & 

{\bf{Competence, relatedness and autonomy}} drive motivation to improve teaching by doing DBER

&
{\bf{Competence and relatedness}} drive motivation to improve teaching by doing DBER; refined conceptualization of short-term plan

\\ \hline
Akemi & 

{\bf{Competence}} drives motivation to improve teaching by doing DBER 
& 
{\bf{Competence and relatedness}} drive motivation to improve teaching by doing DBER; refined conceptualization of short-term plan.

\\ \hline
  \end{tabular}
  \label{tab:selfreactiveness}
\end{table*}

\subsubsection{Pre-PEER Self-reactiveness}
For Olivia, competence and relatedness are the two components of self-determination theory that help her the most in doing DBER at her institution. 
She wants to do MER to increase her competence in teaching to increase student success and persistence in math courses. She also wants to relate her research results from her classrooms to her institution:
\begin{quote}
    \textit{The driving force for me, and I know that math educators often don’t really want to talk about this in this way, has been to see students be more successful. Specifically, to pass at higher rates and to continue sort of to the next course at higher rates. And what, you know, I’m not interested in just what happens in my class, I’m interested in what happens at the institution.}
\end{quote}
To have productive conversations about instructional change in her department, Olivia wants to use the results of her own research-based findings from MER. This refers to the relatedness of doing this type of research as it provides a means to communicate with evidence based, context-specific ways, her research results to her colleagues:
\begin{quote}
    \textit{You know, all my colleagues are math professors, which means you can’t just walk up to them and say, “hey, let’s try this thing.” If you don’t start with something that’s evidence based, if you’re not starting from a point of scholarship, you’re not going to get started.}
\end{quote}
This last excerpt also underlines the value and the potential impact that DBER scholarship can have in bringing many faculty members part of her department on board in making instructional change. Olivia also wants her math department and university to find better ways to assess student learning, which ties into the self-regulation component of self-reactiveness.
She is supported in her DBER engagement because of its potential to addresses critical and current needs at her institution: it is a context-specific, yet research-based way to improve success and retention in mathematics courses.

Similarly competence, relatedness and autonomy are all elements that motivate and self-regulate Madison's engagement in DBER. 
In terms of competence, Madison wants to become a better physics teacher by improving her classroom practices.  She describes doing PER is a way for her to become better at her job:
\begin{quote}
    \textit{I’ve also found myself really interested in physics education research, both as, you know, using it to help inform my teaching, but also, I’m just interested in learning more about how to be a good physics teacher.}
\end{quote}
Features of autonomy and relatedness are discussed when Madison describes the freedom to pursue various new teaching evidence-based strategies in the classroom:

\begin{quote}
\textit{So I feel like there’s been a lot of freedom there to pursue different teaching routes and, you know, this comes up in things like tenure and promotion too. Like, our department puts, I think, a good deal of weight and will give you a lot of credit for going and trying these new pedagogy.} \end{quote}
She articulates that research-based teaching practices are valued in tenure and promotion evaluations. This external regulation provided by her department motivates engagement in instructional change. PER is encouraged due to its potential benefits for student learning in a primarily undergraduate institution that attracts underrepresented groups and wants to best prepare them for their post-undergraduate careers. 

As for Akemi, competence is the most prominent component of self-determination theory that motivates her to engage in DBER. 
She wants to do PER projects to create more equitable learning environments for students in her classroom. She wants to investigate the ways in which she can increase self-efficacy of underrepresented students in her physics courses: 
\begin{quote}
    \textit{I’m interested in that [doing physics education around promoting equity] because I found like minorities in classrooms are usually either lack of self-efficacy, a lack of confidence, or the opposite, they think they are good, they don’t know that they are bad at this stuff. So I just I’m interested in like how my students are doing and how they are thinking, and I think [doing a PER project around] that [can] help.}
\end{quote}
Akemi does not articulate the ways in which her PER work will be evaluated or the ways she will assess her own endeavors in this new field of research for her. This is most likely  due to her being currently in a temporary faculty position during this interview as she says that she is a visiting professor and is not required to do research. 

Pre-PEER in self-reactiveness, we identified that Olivia was motivated to do MER to improve passing and retention rates at her institution. Madison wanted to be a better physics teacher by engaging in PER to improve her classroom teaching practices. Akemi wanted to create more equitable physics classrooms so engaging in PER would allow her to investigate the interplay between self-efficacy and underrepresented populations in physics classrooms. Olivia and Madison related their motivation to the value of doing DBER would bring to convincing colleagues of instructional change, benefiting students at their respective institution and getting recognized for this type work in promotion and tenure evaluations. 

\subsubsection{Post-PEER Self-reactiveness}

For Olivia, competence remains a primary motivator. She continues to want to improve teaching by doing MER at her particular institution.
In terms of self-regulation, Olivia discusses her autonomy as she is reflecting on some of the discussions that occurred at PEER around mentorship and ways one can discuss the value of DBER in a department that may not be supportive of this type of research.  She recognizes how much freedom she has compared to less senior faculty in pursuing DBER: doing a research study, getting results, making recommendations to her department about changes and being heard. 
Although she is already at an institution that values math education research, she is not expecting  rewards from her school to motivate and regulate her engagement in MER. This integrated regulation is considered the type of extrinsic motivation regulator that leads to the most autonomy \cite{ryan2000self}, which shows how Olivia has a high level of autonomy in her MER work. The PEER activities that played a role in her growth in self-reactiveness were topical small group discussions and interactions with a range of career stages. 

Similarly for Madison,  increasing competence remains a major motivator for pursuing PER. Improving teaching practices for the type of students her institution attracts informs her education research interests. 
Through self-reflection on the societal impact of her job, she hopes  to help the student population of her institution get the most out of their education. Improving physics laboratory courses allows the development of technical skills that can be useful for students as they search for jobs after they graduate. The main PEER activity  that played a role in her growth in self-reactiveness was generative writing regularly throughout the workshops. 

Through different programs elements than Olivia and Madison, Akemi refines her interest and her research project’s focus. Her motivation to do PER is still about equity in physics classrooms. By talking to facilitators and engaging with the broader PER literature, she finds ways to specifically enhance her project by situating her work within the vast array of research published about self-efficacy of students in physics classrooms and gender equity \cite{nissen2016gender, kalender2020damage}. Given that she is in-between two temporary teaching positions, she does not elaborate on the ways she will be evaluated in her research endeavors.  The PEER activities that played a role in her growth in self-reactiveness were  interactions with facilitators with DBER expertise and guiding engagement with DBER literature. 

All three case study subjects were consistent in their motivation behind their reason to transition into doing DBER. For Olivia, engagement with participants at PEER further highlighted her freedom to pursue MER at this stage of her career and institution. For Madison, self-reflection on her impact as a physics instructor deepened her motivation to engage in PER to improve learning outcomes for her students. For Akemi, situating her work within PER helped her refine her interest.

\subsection{Self-reflectiveness}
In the literature, self-reflectiveness is defined as belief in one's perceived competence in their ability to undertake a behavior (self-efficacy). In our context,  self-reflectiveness refers to what emerging STEM education researchers perceived competence in DBER to be. 
Similarly to growth in intentionality, forethought and self-reactiveness, growth happened in self-reflectiveness. However, there were more program elements that came into play in this component, making self-reflectiveness the agency component with the most growth. Table \ref{tab:selfreflectivenesss} summarizes the status of self-reflectiveness through pre and post PEER for participants.

\begin{table*}[tbh]
\caption{Status of self-reflectiveness (self-efficacy) pre and post PEER for each participant}
    \begin{tabular}{p{3cm} p{7cm} p{7cm}}
    {\bf{ Participants} }& {\bf Pre-PEER } &{\bf Post-PEER} \\ 
    \hline
    Olivia & 

    {\bf{Low self-efficacy in mastery experiences:}} 

    Looking for mastery experiences

    &
    {\bf{Higher self-efficacy rooted in multiple sources of self-efficacy, especially:}} 

    Vicarious learning and mastery experiences

    \\ \hline
    Madison  & 

    {\bf{Low self-efficacy in mastery experiences:
    }} 

    Looking for mastery experiences

    &
    {\bf{Higher self-efficacy rooted in multiple sources of self-efficacy, especially:}} 

    Physiological state and mastery experiences

    \\ \hline
    Akemi & 

    {\bf{Low self-efficacy in mastery experiences:
    }} 

    Looking for mastery experiences

    & 
    {\bf{Higher self-efficacy rooted in multiple sources of self-efficacy, especially:}} 

    Verbal persuasion and mastery experiences

    \\ \hline
    \end{tabular}
\label{tab:selfreflectivenesss}
\end{table*}

\subsubsection{Pre-PEER Self-reflectiveness}
Olivia expresses low self-efficacy when she describes her NSF grant proposal process she applied for to engage in MER. She did not ask for help from some of her colleagues because she feels unqualified to do MER compared to them despite collaborating with them in other areas of research:
\begin{quote}
    \textit{So the one thing I remember about is my professional colleagues that I didn’t collaborate with. Yeah, so the issue is… So I have, I would say, three colleagues, two of whom I’ve actually written research papers in mathematics with, who have done a lot of math ed grants, actually quite a few. And I did not partner with them [...] I was embarrassed. I know so little about it, even less, you know, and I have to admit, you know, I knew I was doing something for which I was unqualified.}
\end{quote}
Her perception is that she does not have the experience that some other people she knows have when it comes to math education grant writing. Vicarious learning, which emphasizes performance comparison, nurtures this sense of low-self-efficacy.
Mastery experience is another reason for her sense of low self-efficacy. Compared to her math research, she feels unqualified to do math education research because she does not know how to turn all her research interests into research projects, which a task she is confident in doing in her graph theory research.

Similarly, Madison articulates that she lacks confidence in doing PER because she does not know how to carry out the different aspects of the research design.  In terms of mastery experiences, she feels that she lacks competence in carrying out this type of research compared to her ability to do so in her experimental physics work: 
\begin{quote}
    \textit{I think I would really, really love to be more confident in myself for my ability to design and carry out an education research project. Like, especially from the nuts and bolts of the education research side of things.}
\end{quote}
Akemi also articulates low self-efficacy when she describes being unaware how to structure the research process. Even though she has put together a proposal that got accepted and she is already engaging in data collection in her classroom, her confidence in her ability to perform PER is low. She expresses throughout the transcript not knowing how to move forward with different steps of the research process if she gets stuck. 

In self-reflectiveness pre-PEER, there was a common trend of low self-efficacy among all three case study subjects, especially in terms of mastery experiences. 

\subsubsection{Post-PEER Self-reflectiveness}
Olivia feels that she knows more about MER because she engaged with the MER literature and received informational knowledge about procedures of MER at PEER. This addresses mastery experience as she feels she can draw from her rich experience in math research herself to contribute to the field of MER. In turn, she feels more comfortable reaching out to collaborate with others because she has a better sense of what she can do for a project.  Performance and experience comparison with researchers with various backgrounds at PEER, vicarious learning,  contributes to her sense of higher self-efficacy:
\begin{quote}
    \textit{I at least have read some math education research. I at least have gone to a workshop where I learned about some things. I’m not totally ignorant about qualitative research and various kinds of surveys and various things like, you know, getting IRB approval and that sort of thing. I’m not just a complete dead weight to someone else who’s doing DBER research, if that makes sense. I don’t want to be dead weight}
\end{quote}
Post-PEER, her confidence level is higher. She articulates the ways she feels that she can bring something useful to MER and serve as a bridging role among communities. She feels more confident in engaging with the MER community. The PEER activities that played a role in her growth in self-reflectiveness were interactions with facilitators with DBER expertise, interactions with a range of career stages,  guided engagement with DBER literature and procedural knowledge workshops. 

For Madison, increase in self-efficacy is seen through the way she describes the impact of receiving concrete get-started information about PER.  Narrowing down her research project to specific steps to write a grant proposal has helped her increase her sense of mastery experience, in turn her self-efficacy:
\begin{quote}
    \textit{So one of my big goals from this whole thing was like to feel confident enough that I could write a grant for my project. And I feel comfortable, much more comfortable now, that I could put a grant together because I have a much better awareness of like the literature I should be looking for and stuff like that. So that was really nice. Yeah, for me, a lot of the skills are the like what do you… Like, I am much more confident in my ability to start a project.}
\end{quote}
Spending time articulating her research interest into a research question has also contributed to her gain confidence in the work she is doing, addressing the physiological component of self-efficacy:
\begin{quote}
    \textit{I feel really, really good and confident that I came out with some idea on okay, how do I go from it’s something I might be interested in to carving that into a research question and start to get the research done.}
\end{quote}
The PEER activities that played a role in her growth in self-reflectiveness were guided engagement with DBER literature, topical small group discussions, procedural knowledge workshops and taking the time to do project mapping of goals with specific tasks for both the near (days, weeks) and far (months) future.

For Akemi, it was comforting to get feedback on the work she was doing and how it may be useful to the DBER community. Given that she is new to the field, she feared that she might have missed someone else's publication. Verbal persuasion, which occurs when Akemi gets real time constructive feedback from peers and facilitators, plays a positive role in increasing her self-efficacy:
\begin{quote}
    \textit{So I kind of asked them whether I should… So I say, I’ve already my project on how oral quizzes impact students’ self-efficacy, and then they told me oh, it’s an interesting project, and then it’s not been done. So I think that’s very important information because I’m new to the education, to this field, and although I’ve already did the literature survey and did not find something similar, I always worry like whether I’ve missed some publication}
\end{quote}
Verbal persuasion occurs when Akemi discusses how supportive the feedback at PEER was. She feels that people were enthusiastic about and valued her ideas: 
\begin{quote}
    \textit{I feel like, I don’t know, I don’t feel this often, I hadn’t felt that my opinions are valuable in research for years. At least that’s not my general feeling in my PhD research, so when I kind of talk [...] they[facilitators] really value what I said, and I think that really boosts my self confidence in this area, like feel I can do educational research like that. So I think that helps a lot.}
\end{quote}
Feeling valued in her research endeavors is an element of PEER that boosted Akemi’s self-confidence because she had not felt it in her condensed matter research during her graduate studies. 
Engaging in regular generative writing really helped her increase her sense of competence, mastery experience:
\begin{quote}
\textit{I’m looking at my project, “okay, I can write something out of it,” and then the generative writing sections are really helpful. I don’t know how to do that at the beginning, it’s painful to write, I really hate writing. But now I can really sit down, wow, I can keep typing for one hour, or like half hour. It’s something that I could not imagine me doing, so I think there’s definitely some change in my ability to move my project forward.}
\end{quote}
The PEER activities that played a role in her growth in self-reflectiveness were interactions with a range of career stages, individualized project feedback from facilitators and generative writing. 

For Olivia, engagement with PEER participants and facilitators, specifically discussing similar interests and comparing experiences with others at different stages of their DBER projects, increased her self-efficacy. For Madison, getting informational knowledge and turning her research interest into a research question translated into gain in self-efficacy. For Akemi, supportive real-time constructive feedback allowed to situate herself within the field and feel welcomed in this field of research, which boosted her self-efficacy.

\section{Discussion}
The key results of our analysis are summarized in Table \ref{tab:discussion} where we highlight program activities as affecting participants' agency, within the theoretical framework. In the table, project mapping refers to mapping of goals with specific tasks for both the near (days, weeks) and far (months) future. Topical discussions refers to topical small group discussions. Facilitator interactions refer to interactions with facilitators with DBER expertise. Career-stage interactions refer to interactions with participants in range of career stages. Project feedback refers to individualized project feedback from facilitators. Generative writing refers to writing as a generative process to keep track of research process, ideas and next steps. DBER literature refers to guiding engagement with DBER literature. Procedural knowledge refers to procedural knowledge workshops.

\begin{table*}[tbh]
\caption{Case study participants highlight program activities affecting their agency, identified within the theoretical framework. }
  \begin{tabular}{p{3.8cm} p{4cm} p{4cm} p{4cm}}
    \hline
{\bf{Theory}}  & Olivia & Madison & Akemi  \\ \hline
{\bf Intentionality } &  Project mapping  & Project mapping  & Project mapping 
\\ \hline
{\bf Forethought} & \raggedright{Facilitator} interactions 

Topical discussions

DBER literature 
& Topical discussions

Project feedback 

Procedural knowledge 

&
DBER literature

Procedural knowledge

\\ \hline
{\bf Self-reactiveness}  & Topical discussions 

\raggedright{Career-stage interactions}

&Generative writing

&Facilitator 

interactions

DBER literature \\ \hline
{\bf Self-reflectiveness} & \raggedright{Facilitator interactions} 

Career- stage interactions 

DBER literature 

Procedural knowledge & \raggedright{Topical discussions} 
\raggedright{DBER literature} 
Procedural knowledge 
Project mapping & Career-stage interactions

Project feedback

Generative writing
\\ \hline 
  \end{tabular}
  \label{tab:discussion}
\end{table*}


\subsection{Interactions within participants for each aspect of agency growth}
Faculty professional development is highly dependent on home institution type, department priorities, and faculty career stage. As such, to understand how participants develop their agency in this new area of research, it is interesting to see how agency components evolve depending on each participant particular career stage and context.

As a Full Professor in a math department, in the course of her engagement with PEER, Olivia focused on developing her existing research project and developed an understanding of where she can situate herself in mathematics education research (MER). The  most noticeable growth in agency occurred thanks to her engagement with participants at various career stages and with various DBER expertise. This engagement really highlighted the autonomy she has as a Full Professor in her research endeavors in MER,  leading to growth in self-reactiveness. This also  translated into growth in self-efficacy as she was able to articulate what she could contribute to the field when engaging with both the math and math education research communities. 

As an Associate Professor in a physics department, Madison focused on narrowing down her research questions and getting started on writing an NSF grant to fund her physics education research (PER) project to expand her research portfolio. As a tenured professor, she has some leeway in pursuing different research interests, especially when finding evidence-based practices contextualized in her department is increasingly becoming a priority for her institutions. The procedural knowledge and the time to reflect and articulate her research interest during PEER led to growth in forethought and self-reflectiveness, leading to overall gain in agency. 

As an early-career professional, in the course of her engagement with PEER,  Akemi focused on refining her research project and getting started with data collection for her project to see where PER could fit within her career trajectory, which led to overall growth in the agency. Mentorship and guidance from PEER facilitators, increased her sense of competence in self-reflectiveness and refined her motivation in self-reactiveness to pursue her research projects in PER.

Although we see agency  growth for each participant in this study, this exploratory analysis draws upon self-reported data of three faculty’s experiences, which cannot be generalized to all emerging STEM education researchers.  Future work should include other participants’ experiences to explore contrasting experiences with agency growth, especially for STEM faculty at different career stages and at different types of institutions. 

\subsection{Program activities across theory elements}
Exploring the impact of program activities across agency components provides evidence of activities that impact agency when designing a professional development program.

Supporting activities in the  growth of self-reactiveness were discussions of similar interest with participants and facilitators, engagement with key DBER literature and opportunities for self-reflection.  Growth in intentionality occurred through the setting of specific  DBER plans for the future, which enabled participants to break down research projects into specific and measurable steps to move forward. Growth in forethought occurred through receiving get-started information, engagement with peers, engagement with the DBER literature and the division of tasks into manageable pieces with multiple iterations. All these elements provided participants the opportunity to refine their projects and anticipate specific actions and consequences they foresee as they move forward with their projects. 

One or a combination of sources of self-efficacy contributed to growth in self-reflectiveness. Verbal persuasion learning through getting real time constructive feedback translated to self-efficacy increase. Vicarious learning through comparison of researchers' expertise with various backgrounds contributed to an increase in self-efficacy. Mastery experiences occurred through transformation of general interest to specific questions and receiving procedural knowledge about the field.  Articulation of realistic and specific addressed the physiological component of self-efficacy. 

Bandura says that self-efficacy is one of the strongest components in agency growth during change and adaptation in the workplace \cite{bandura_functional_2012}. It is not surprising that increased self-efficacy echoed more broadly to gain in other areas of the agency framework. Nonetheless, varying and overlapping activities resonated with participants, which showcase various possible ways a professional development can contribute to increase a sense of agency in a new research area. 

Program elements discussed in self-reflectiveness are the only ones that span across all other components of agency (forethought, intentionality, self-reactiveness). Our case study participants  articulated the ways in which built-in within the structure of PEER are activities and interactions that address each component of self-efficacy. These elements of PEER that increase self-efficacy carry over to the three other components of agency, leading to overall gain in agency. Program elements in forethought, intentionality and self-reactiveness stem from any exposure to the research process. They are not unique to getting started in DBER, exposure and engagement with a research community will inevitably refine ideas in each of those areas. However, what we find is that the PEER program provides structure to these elements that seem to resonate quite strongly with participants. PEER provides the space and community to be an agentic emerging STEM education researcher. PEER  facilitates engagement in research tasks that jump start emerging STEM education research transition into DBER, especially when they have extensive training in other areas of research and experience in teaching. Thus, STEM faculty who already have extensive training in research and myriad of teaching experiences in their specific discipline can chair their first research project in DBER when agency is a central tenet of the professional development opportunities they engage in. 

In contrast, some program elements such as the setting of expectations norms and some procedural workshops (e.g. observational data and theory workshops)  were not brought by these three case study participants. In this analysis, they were not factors explicitly affecting  their agency. However, this does not mean that these activities do not affect other participants’ agency and/or  have a programmatic impact that leads to agency growth. First, the  expectations and norms setting puts forward the principles of PEER for participants engagement and community building, which makes this professional development opportunity an experiential learning experience in which agency growth happens as a consequence of that. Second, the specific workshops not brought up may not have impacted agency development for  these participants, but may have done so on  others depending on where they are at with their research. If their research interest is not immediately tied with observational data, it may not have had a significant enough impact to be brought up during interviews. In addition, we are looking at growth and some topics such as theory that are overwhelming and an area of struggle for emerging education research \cite{hass2022emerging} may not be brought up in this analysis lens.

DBER's interdisciplinarity and the myriad of ways it is conducted can be challenging for new researchers interested in the field. For emerging STEM education researchers, finding professional development that addresses their concerns from an agentic perspective is a need that must be fulfilled. Support structures can come in various ways, but our research shows the process by which a professional development opportunity worked in favor of increasing self-efficacy and echoed more broadly into agency. This agency growth can sustain engagement in DBER and increase DBER research in different institutional contexts and improve STEM education through effective evidence-based practices that stem from the particular needs of the institutional contexts in which the research interest orignates. To build capacity and community for STEM education research, the DBER community should create professional development opportunities that focus on supporting agency in engaging in DBER, particularly self-efficacy, for STEM emerging education researchers.

\section{Conclusion}
To improve STEM education, some STEM faculty jump start their transition in DBER at different stages of their career. To support their endeavors to conduct DBER in different instructional settings, our study identified elements of a professional development program that increase agency.  Our case study analysis showed that addressing one or a combination of self-efficacy sources echoed into growth of other components of agency. This overall gain of agency supports emerging discipline-based education researchers' transition to the field.  

\begin{acknowledgments}
We thank Christopher Hass for his help with the data collection and inter-rater reliability. We thank Elizabeth Kustusch for transcribing our interviews. We also thank PEER participants, particularly our case study participants, for allowing us to study the experiences of emerging STEM education researchers. We also would like to thank the PEER facilitators who contributed to this virtual edition of the program. This work was supported by NSF DUE 2025174 / 2025170 and 1726479 / 1726113.
\end{acknowledgments}

\bibliography{apssamp}

\begin{thebibliography}{47}%
\makeatletter
\providecommand \@ifxundefined [1]{%
 \@ifx{#1\undefined}
}%
\providecommand \@ifnum [1]{%
 \ifnum #1\expandafter \@firstoftwo
 \else \expandafter \@secondoftwo
 \fi
}%
\providecommand \@ifx [1]{%
 \ifx #1\expandafter \@firstoftwo
 \else \expandafter \@secondoftwo
 \fi
}%
\providecommand \natexlab [1]{#1}%
\providecommand \enquote  [1]{``#1''}%
\providecommand \bibnamefont  [1]{#1}%
\providecommand \bibfnamefont [1]{#1}%
\providecommand \citenamefont [1]{#1}%
\providecommand \href@noop [0]{\@secondoftwo}%
\providecommand \href [0]{\begingroup \@sanitize@url \@href}%
\providecommand \@href[1]{\@@startlink{#1}\@@href}%
\providecommand \@@href[1]{\endgroup#1\@@endlink}%
\providecommand \@sanitize@url [0]{\catcode `\\12\catcode `\$12\catcode
  `\&12\catcode `\#12\catcode `\^12\catcode `\_12\catcode `\%12\relax}%
\providecommand \@@startlink[1]{}%
\providecommand \@@endlink[0]{}%
\providecommand \url  [0]{\begingroup\@sanitize@url \@url }%
\providecommand \@url [1]{\endgroup\@href {#1}{\urlprefix }}%
\providecommand \urlprefix  [0]{URL }%
\providecommand \Eprint [0]{\href }%
\providecommand \doibase [0]{https://doi.org/}%
\providecommand \selectlanguage [0]{\@gobble}%
\providecommand \bibinfo  [0]{\@secondoftwo}%
\providecommand \bibfield  [0]{\@secondoftwo}%
\providecommand \translation [1]{[#1]}%
\providecommand \BibitemOpen [0]{}%
\providecommand \bibitemStop [0]{}%
\providecommand \bibitemNoStop [0]{.\EOS\space}%
\providecommand \EOS [0]{\spacefactor3000\relax}%
\providecommand \BibitemShut  [1]{\csname bibitem#1\endcsname}%
\let\auto@bib@innerbib\@empty
\bibitem [{\citenamefont {Council}\ \emph {et~al.}(2012)\citenamefont {Council}
  \emph {et~al.}}]{national2012discipline}%
  \BibitemOpen
  \bibfield  {author} {\bibinfo {author} {\bibfnamefont {N.~R.}\ \bibnamefont
  {Council}} \emph {et~al.},\ }\bibfield  {title} {\bibinfo {title}
  {Discipline-based education research: Understanding and improving learning in
  undergraduate science and engineering},\ }\href@noop {} {\bibfield  {journal}
  {\bibinfo  {journal} {National Academies Press}\ } (\bibinfo {year}
  {2012})}\BibitemShut {NoStop}%
\bibitem [{\citenamefont {APS}(2019)}]{aps992}%
  \BibitemOpen
  \bibfield  {author} {\bibinfo {author} {\bibnamefont {APS}},\ }\bibfield
  {title} {\bibinfo {title} {Statement 99.2 research in physics education},\
  }\href@noop {} {\bibfield  {journal} {\bibinfo  {journal} {American Physical
  Society}\ } (\bibinfo {year} {2019})}\BibitemShut {NoStop}%
\bibitem [{\citenamefont {Barthelemy}\ \emph {et~al.}(2013)\citenamefont
  {Barthelemy}, \citenamefont {Henderson},\ and\ \citenamefont
  {Grunert}}]{barthelemy2013they}%
  \BibitemOpen
  \bibfield  {author} {\bibinfo {author} {\bibfnamefont {R.~S.}\ \bibnamefont
  {Barthelemy}}, \bibinfo {author} {\bibfnamefont {C.}~\bibnamefont
  {Henderson}},\ and\ \bibinfo {author} {\bibfnamefont {M.~L.}\ \bibnamefont
  {Grunert}},\ }\bibfield  {title} {\bibinfo {title} {How do they get here?:
  Paths into physics education research},\ }\href@noop {} {\bibfield  {journal}
  {\bibinfo  {journal} {Physical Review Special Topics-Physics Education
  Research}\ }\textbf {\bibinfo {volume} {9}},\ \bibinfo {pages} {020107}
  (\bibinfo {year} {2013})}\BibitemShut {NoStop}%
\bibitem [{\citenamefont {Van~Dusen}\ \emph {et~al.}(2014)\citenamefont
  {Van~Dusen}, \citenamefont {Barthelemy},\ and\ \citenamefont
  {Henderson}}]{van2014educational}%
  \BibitemOpen
  \bibfield  {author} {\bibinfo {author} {\bibfnamefont {B.}~\bibnamefont
  {Van~Dusen}}, \bibinfo {author} {\bibfnamefont {R.~S.}\ \bibnamefont
  {Barthelemy}},\ and\ \bibinfo {author} {\bibfnamefont {C.}~\bibnamefont
  {Henderson}},\ }\bibfield  {title} {\bibinfo {title} {Educational
  trajectories of graduate students in physics education research},\
  }\href@noop {} {\bibfield  {journal} {\bibinfo  {journal} {Physical Review
  Special Topics-Physics Education Research}\ }\textbf {\bibinfo {volume}
  {10}},\ \bibinfo {pages} {020106} (\bibinfo {year} {2014})}\BibitemShut
  {NoStop}%
\bibitem [{\citenamefont {Anderson}\ \emph {et~al.}(2017)\citenamefont
  {Anderson}, \citenamefont {Crespi},\ and\ \citenamefont
  {Sayre}}]{anderson2017linking}%
  \BibitemOpen
  \bibfield  {author} {\bibinfo {author} {\bibfnamefont {K.~A.}\ \bibnamefont
  {Anderson}}, \bibinfo {author} {\bibfnamefont {M.}~\bibnamefont {Crespi}},\
  and\ \bibinfo {author} {\bibfnamefont {E.~C.}\ \bibnamefont {Sayre}},\
  }\bibfield  {title} {\bibinfo {title} {Linking behavior in the physics
  education research coauthorship network},\ }\href@noop {} {\bibfield
  {journal} {\bibinfo  {journal} {Physical Review Physics Education Research}\
  }\textbf {\bibinfo {volume} {13}},\ \bibinfo {pages} {010121} (\bibinfo
  {year} {2017})}\BibitemShut {NoStop}%
\bibitem [{\citenamefont {Bush}\ \emph {et~al.}(2011)\citenamefont {Bush},
  \citenamefont {Pelaez}, \citenamefont {Rudd}, \citenamefont {Stevens},
  \citenamefont {Tanner},\ and\ \citenamefont
  {Williams}}]{bush2011investigation}%
  \BibitemOpen
  \bibfield  {author} {\bibinfo {author} {\bibfnamefont {S.~D.}\ \bibnamefont
  {Bush}}, \bibinfo {author} {\bibfnamefont {N.~J.}\ \bibnamefont {Pelaez}},
  \bibinfo {author} {\bibfnamefont {J.~A.}\ \bibnamefont {Rudd}}, \bibinfo
  {author} {\bibfnamefont {M.~T.}\ \bibnamefont {Stevens}}, \bibinfo {author}
  {\bibfnamefont {K.~D.}\ \bibnamefont {Tanner}},\ and\ \bibinfo {author}
  {\bibfnamefont {K.~S.}\ \bibnamefont {Williams}},\ }\bibfield  {title}
  {\bibinfo {title} {Investigation of science faculty with education
  specialties within the largest university system in the united states},\
  }\href@noop {} {\bibfield  {journal} {\bibinfo  {journal} {CBE—Life
  Sciences Education}\ }\textbf {\bibinfo {volume} {10}},\ \bibinfo {pages}
  {25} (\bibinfo {year} {2011})}\BibitemShut {NoStop}%
\bibitem [{\citenamefont {Bush}\ \emph {et~al.}(2013)\citenamefont {Bush},
  \citenamefont {Pelaez}, \citenamefont {Rudd}, \citenamefont {Stevens},
  \citenamefont {Tanner},\ and\ \citenamefont {Williams}}]{bush2013widespread}%
  \BibitemOpen
  \bibfield  {author} {\bibinfo {author} {\bibfnamefont {S.~D.}\ \bibnamefont
  {Bush}}, \bibinfo {author} {\bibfnamefont {N.~J.}\ \bibnamefont {Pelaez}},
  \bibinfo {author} {\bibfnamefont {J.~A.}\ \bibnamefont {Rudd}}, \bibinfo
  {author} {\bibfnamefont {M.~T.}\ \bibnamefont {Stevens}}, \bibinfo {author}
  {\bibfnamefont {K.~D.}\ \bibnamefont {Tanner}},\ and\ \bibinfo {author}
  {\bibfnamefont {K.~S.}\ \bibnamefont {Williams}},\ }\bibfield  {title}
  {\bibinfo {title} {Widespread distribution and unexpected variation among
  science faculty with education specialties (sfes) across the united states},\
  }\href@noop {} {\bibfield  {journal} {\bibinfo  {journal} {Proceedings of the
  National Academy of Sciences}\ }\textbf {\bibinfo {volume} {110}},\ \bibinfo
  {pages} {7170} (\bibinfo {year} {2013})}\BibitemShut {NoStop}%
\bibitem [{\citenamefont {Bush}\ \emph {et~al.}(2016)\citenamefont {Bush},
  \citenamefont {Rudd}, \citenamefont {Stevens}, \citenamefont {Tanner},\ and\
  \citenamefont {Williams}}]{bush2016fostering}%
  \BibitemOpen
  \bibfield  {author} {\bibinfo {author} {\bibfnamefont {S.~D.}\ \bibnamefont
  {Bush}}, \bibinfo {author} {\bibfnamefont {J.~A.}\ \bibnamefont {Rudd}},
  \bibinfo {author} {\bibfnamefont {M.~T.}\ \bibnamefont {Stevens}}, \bibinfo
  {author} {\bibfnamefont {K.~D.}\ \bibnamefont {Tanner}},\ and\ \bibinfo
  {author} {\bibfnamefont {K.~S.}\ \bibnamefont {Williams}},\ }\bibfield
  {title} {\bibinfo {title} {Fostering change from within: Influencing teaching
  practices of departmental colleagues by science faculty with education
  specialties},\ }\href@noop {} {\bibfield  {journal} {\bibinfo  {journal}
  {PloS one}\ }\textbf {\bibinfo {volume} {11}},\ \bibinfo {pages} {e0150914}
  (\bibinfo {year} {2016})}\BibitemShut {NoStop}%
\bibitem [{\citenamefont {Bush}\ \emph {et~al.}(2017)\citenamefont {Bush},
  \citenamefont {Stevens}, \citenamefont {Tanner},\ and\ \citenamefont
  {Williams}}]{bush2017origins}%
  \BibitemOpen
  \bibfield  {author} {\bibinfo {author} {\bibfnamefont {S.~D.}\ \bibnamefont
  {Bush}}, \bibinfo {author} {\bibfnamefont {M.~T.}\ \bibnamefont {Stevens}},
  \bibinfo {author} {\bibfnamefont {K.~D.}\ \bibnamefont {Tanner}},\ and\
  \bibinfo {author} {\bibfnamefont {K.~S.}\ \bibnamefont {Williams}},\
  }\bibfield  {title} {\bibinfo {title} {Origins of science faculty with
  education specialties: Hiring motivations and prior connections explain
  institutional differences in the sfes phenomenon},\ }\href@noop {} {\bibfield
   {journal} {\bibinfo  {journal} {BioScience}\ }\textbf {\bibinfo {volume}
  {67}},\ \bibinfo {pages} {452} (\bibinfo {year} {2017})}\BibitemShut
  {NoStop}%
\bibitem [{\citenamefont {Bush}\ \emph {et~al.}(2020)\citenamefont {Bush},
  \citenamefont {Stevens}, \citenamefont {Tanner},\ and\ \citenamefont
  {Williams}}]{bush2020disciplinary}%
  \BibitemOpen
  \bibfield  {author} {\bibinfo {author} {\bibfnamefont {S.~D.}\ \bibnamefont
  {Bush}}, \bibinfo {author} {\bibfnamefont {M.~T.}\ \bibnamefont {Stevens}},
  \bibinfo {author} {\bibfnamefont {K.~D.}\ \bibnamefont {Tanner}},\ and\
  \bibinfo {author} {\bibfnamefont {K.~S.}\ \bibnamefont {Williams}},\
  }\bibfield  {title} {\bibinfo {title} {Disciplinary bias, money matters, and
  persistence: Deans’ perspectives on science faculty with education
  specialties (sfes)},\ }\href@noop {} {\bibfield  {journal} {\bibinfo
  {journal} {CBE—Life Sciences Education}\ }\textbf {\bibinfo {volume}
  {19}},\ \bibinfo {pages} {ar34} (\bibinfo {year} {2020})}\BibitemShut
  {NoStop}%
\bibitem [{\citenamefont {Shipley}\ \emph {et~al.}(2017)\citenamefont
  {Shipley}, \citenamefont {McConnell}, \citenamefont {McNeal}, \citenamefont
  {Petcovic},\ and\ \citenamefont {John}}]{shipley2017transdisciplinary}%
  \BibitemOpen
  \bibfield  {author} {\bibinfo {author} {\bibfnamefont {T.~F.}\ \bibnamefont
  {Shipley}}, \bibinfo {author} {\bibfnamefont {D.}~\bibnamefont {McConnell}},
  \bibinfo {author} {\bibfnamefont {K.~S.}\ \bibnamefont {McNeal}}, \bibinfo
  {author} {\bibfnamefont {H.~L.}\ \bibnamefont {Petcovic}},\ and\ \bibinfo
  {author} {\bibfnamefont {K.~E.~S.}\ \bibnamefont {John}},\ }\bibfield
  {title} {\bibinfo {title} {Transdisciplinary science education research and
  practice: Opportunities for ger in a developing stem discipline-based
  education research alliance (dber-a)},\ }\href@noop {} {\bibfield  {journal}
  {\bibinfo  {journal} {Journal of Geoscience Education}\ }\textbf {\bibinfo
  {volume} {65}},\ \bibinfo {pages} {354} (\bibinfo {year} {2017})}\BibitemShut
  {NoStop}%
\bibitem [{\citenamefont {Bush}\ \emph {et~al.}(2019)\citenamefont {Bush},
  \citenamefont {Stevens}, \citenamefont {Tanner},\ and\ \citenamefont
  {Williams}}]{bush2019evolving}%
  \BibitemOpen
  \bibfield  {author} {\bibinfo {author} {\bibfnamefont {S.~D.}\ \bibnamefont
  {Bush}}, \bibinfo {author} {\bibfnamefont {M.~T.}\ \bibnamefont {Stevens}},
  \bibinfo {author} {\bibfnamefont {K.~D.}\ \bibnamefont {Tanner}},\ and\
  \bibinfo {author} {\bibfnamefont {K.~S.}\ \bibnamefont {Williams}},\
  }\bibfield  {title} {\bibinfo {title} {Evolving roles of scientists as change
  agents in science education over a decade: Sfes roles beyond discipline-based
  education research},\ }\href@noop {} {\bibfield  {journal} {\bibinfo
  {journal} {Science advances}\ }\textbf {\bibinfo {volume} {5}},\ \bibinfo
  {pages} {eaav6403} (\bibinfo {year} {2019})}\BibitemShut {NoStop}%
\bibitem [{\citenamefont {Franklin}\ \emph {et~al.}(2018)\citenamefont
  {Franklin}, \citenamefont {Sayre},\ and\ \citenamefont
  {Kustusch}}]{franklin2018peer}%
  \BibitemOpen
  \bibfield  {author} {\bibinfo {author} {\bibfnamefont {S.}~\bibnamefont
  {Franklin}}, \bibinfo {author} {\bibfnamefont {E.~C.}\ \bibnamefont
  {Sayre}},\ and\ \bibinfo {author} {\bibfnamefont {M.~B.}\ \bibnamefont
  {Kustusch}},\ }\bibfield  {title} {\bibinfo {title} {Peer: Professional
  development experiences for education researchers},\ }\href@noop {}
  {\bibfield  {journal} {\bibinfo  {journal} {ASEE Annual Conference and
  Exposition Conference Proceedings}\ } (\bibinfo {year} {2018})}\BibitemShut
  {NoStop}%
\bibitem [{\citenamefont {Hass}\ \emph {et~al.}(2021)\citenamefont {Hass},
  \citenamefont {Hancock}, \citenamefont {Wilson}, \citenamefont {El-Adawy},\
  and\ \citenamefont {Sayre}}]{hass2021community}%
  \BibitemOpen
  \bibfield  {author} {\bibinfo {author} {\bibfnamefont {C.~A.}\ \bibnamefont
  {Hass}}, \bibinfo {author} {\bibfnamefont {E.}~\bibnamefont {Hancock}},
  \bibinfo {author} {\bibfnamefont {S.}~\bibnamefont {Wilson}}, \bibinfo
  {author} {\bibfnamefont {S.}~\bibnamefont {El-Adawy}},\ and\ \bibinfo
  {author} {\bibfnamefont {E.~C.}\ \bibnamefont {Sayre}},\ }\bibfield  {title}
  {\bibinfo {title} {Community roles for supporting emerging education
  researchers},\ }\href@noop {} {\bibfield  {journal} {\bibinfo  {journal}
  {Physics Education Research Conference Proceedings}\ ,\ \bibinfo {pages}
  {172}} (\bibinfo {year} {2021})}\BibitemShut {NoStop}%
\bibitem [{\citenamefont {Daniel}\ \emph {et~al.}(2022)\citenamefont {Daniel},
  \citenamefont {McConnell}, \citenamefont {Schuchardt},\ and\ \citenamefont
  {Peffer}}]{daniel2022challenges}%
  \BibitemOpen
  \bibfield  {author} {\bibinfo {author} {\bibfnamefont {K.~L.}\ \bibnamefont
  {Daniel}}, \bibinfo {author} {\bibfnamefont {M.}~\bibnamefont {McConnell}},
  \bibinfo {author} {\bibfnamefont {A.}~\bibnamefont {Schuchardt}},\ and\
  \bibinfo {author} {\bibfnamefont {M.~E.}\ \bibnamefont {Peffer}},\ }\bibfield
   {title} {\bibinfo {title} {Challenges facing interdisciplinary researchers:
  Findings from a professional development workshop},\ }\href@noop {}
  {\bibfield  {journal} {\bibinfo  {journal} {Plos one}\ }\textbf {\bibinfo
  {volume} {17}},\ \bibinfo {pages} {e0267234} (\bibinfo {year}
  {2022})}\BibitemShut {NoStop}%
\bibitem [{\citenamefont {El-Adawy}\ \emph {et~al.}(2022)\citenamefont
  {El-Adawy}, \citenamefont {Huynh}, \citenamefont {Kustusch},\ and\
  \citenamefont {Sayre}}]{el2022context}%
  \BibitemOpen
  \bibfield  {author} {\bibinfo {author} {\bibfnamefont {S.}~\bibnamefont
  {El-Adawy}}, \bibinfo {author} {\bibfnamefont {T.}~\bibnamefont {Huynh}},
  \bibinfo {author} {\bibfnamefont {M.~B.}\ \bibnamefont {Kustusch}},\ and\
  \bibinfo {author} {\bibfnamefont {E.~C.}\ \bibnamefont {Sayre}},\ }\bibfield
  {title} {\bibinfo {title} {Context interactions and physics faculty’s
  professional development: Case study},\ }\href@noop {} {\bibfield  {journal}
  {\bibinfo  {journal} {Physical Review Physics Education Research}\ }\textbf
  {\bibinfo {volume} {18}},\ \bibinfo {pages} {020104} (\bibinfo {year}
  {2022})}\BibitemShut {NoStop}%
\bibitem [{\citenamefont {Strubbe}\ \emph {et~al.}(2020)\citenamefont
  {Strubbe}, \citenamefont {Madsen}, \citenamefont {McKagan},\ and\
  \citenamefont {Sayre}}]{strubbe2020beyond}%
  \BibitemOpen
  \bibfield  {author} {\bibinfo {author} {\bibfnamefont {L.~E.}\ \bibnamefont
  {Strubbe}}, \bibinfo {author} {\bibfnamefont {A.~M.}\ \bibnamefont {Madsen}},
  \bibinfo {author} {\bibfnamefont {S.~B.}\ \bibnamefont {McKagan}},\ and\
  \bibinfo {author} {\bibfnamefont {E.~C.}\ \bibnamefont {Sayre}},\ }\bibfield
  {title} {\bibinfo {title} {Beyond teaching methods: Highlighting physics
  faculty’s strengths and agency},\ }\href@noop {} {\bibfield  {journal}
  {\bibinfo  {journal} {Physical Review Physics Education Research}\ }\textbf
  {\bibinfo {volume} {16}},\ \bibinfo {pages} {020105} (\bibinfo {year}
  {2020})}\BibitemShut {NoStop}%
\bibitem [{\citenamefont {Gonzales}(2015)}]{gonzales2015faculty}%
  \BibitemOpen
  \bibfield  {author} {\bibinfo {author} {\bibfnamefont {L.~D.}\ \bibnamefont
  {Gonzales}},\ }\bibfield  {title} {\bibinfo {title} {Faculty agency in
  striving university contexts: Mundane yet powerful acts of agency},\
  }\href@noop {} {\bibfield  {journal} {\bibinfo  {journal} {British
  Educational Research Journal}\ }\textbf {\bibinfo {volume} {41}},\ \bibinfo
  {pages} {303} (\bibinfo {year} {2015})}\BibitemShut {NoStop}%
\bibitem [{\citenamefont {Du}\ \emph {et~al.}(2021)\citenamefont {Du},
  \citenamefont {Naji}, \citenamefont {Ebead},\ and\ \citenamefont
  {Ma}}]{du2021engineering}%
  \BibitemOpen
  \bibfield  {author} {\bibinfo {author} {\bibfnamefont {X.}~\bibnamefont
  {Du}}, \bibinfo {author} {\bibfnamefont {K.~K.}\ \bibnamefont {Naji}},
  \bibinfo {author} {\bibfnamefont {U.}~\bibnamefont {Ebead}},\ and\ \bibinfo
  {author} {\bibfnamefont {J.}~\bibnamefont {Ma}},\ }\bibfield  {title}
  {\bibinfo {title} {Engineering instructors’ professional agency development
  and identity renegotiation through engaging in pedagogical change towards
  pbl},\ }\href@noop {} {\bibfield  {journal} {\bibinfo  {journal} {European
  Journal of Engineering Education}\ }\textbf {\bibinfo {volume} {46}},\
  \bibinfo {pages} {116} (\bibinfo {year} {2021})}\BibitemShut {NoStop}%
\bibitem [{\citenamefont {Lande}\ and\ \citenamefont
  {Mesa}(2016)}]{lande2016instructional}%
  \BibitemOpen
  \bibfield  {author} {\bibinfo {author} {\bibfnamefont {E.}~\bibnamefont
  {Lande}}\ and\ \bibinfo {author} {\bibfnamefont {V.}~\bibnamefont {Mesa}},\
  }\bibfield  {title} {\bibinfo {title} {Instructional decision making and
  agency of community college mathematics faculty},\ }\href@noop {} {\bibfield
  {journal} {\bibinfo  {journal} {ZDM}\ }\textbf {\bibinfo {volume} {48}},\
  \bibinfo {pages} {199} (\bibinfo {year} {2016})}\BibitemShut {NoStop}%
\bibitem [{\citenamefont {Ndunda}\ \emph {et~al.}(2017)\citenamefont {Ndunda},
  \citenamefont {Van~Sickle}, \citenamefont {Perry},\ and\ \citenamefont
  {Capelloni}}]{ndunda2017university}%
  \BibitemOpen
  \bibfield  {author} {\bibinfo {author} {\bibfnamefont {M.}~\bibnamefont
  {Ndunda}}, \bibinfo {author} {\bibfnamefont {M.}~\bibnamefont {Van~Sickle}},
  \bibinfo {author} {\bibfnamefont {L.}~\bibnamefont {Perry}},\ and\ \bibinfo
  {author} {\bibfnamefont {A.}~\bibnamefont {Capelloni}},\ }\bibfield  {title}
  {\bibinfo {title} {University- urban high school partnership: Math and
  science professional learning communities},\ }\href@noop {} {\bibfield
  {journal} {\bibinfo  {journal} {School Science and Mathematics}\ }\textbf
  {\bibinfo {volume} {117}},\ \bibinfo {pages} {137} (\bibinfo {year}
  {2017})}\BibitemShut {NoStop}%
\bibitem [{\citenamefont {Quan}\ \emph {et~al.}(2019)\citenamefont {Quan},
  \citenamefont {Corbo}, \citenamefont {Finkelstein}, \citenamefont {Pawlak},
  \citenamefont {Falkenberg}, \citenamefont {Geanious}, \citenamefont {Ngai},
  \citenamefont {Smith}, \citenamefont {Wise}, \citenamefont {Pilgrim} \emph
  {et~al.}}]{quan2019designing}%
  \BibitemOpen
  \bibfield  {author} {\bibinfo {author} {\bibfnamefont {G.~M.}\ \bibnamefont
  {Quan}}, \bibinfo {author} {\bibfnamefont {J.~C.}\ \bibnamefont {Corbo}},
  \bibinfo {author} {\bibfnamefont {N.~D.}\ \bibnamefont {Finkelstein}},
  \bibinfo {author} {\bibfnamefont {A.}~\bibnamefont {Pawlak}}, \bibinfo
  {author} {\bibfnamefont {K.}~\bibnamefont {Falkenberg}}, \bibinfo {author}
  {\bibfnamefont {C.}~\bibnamefont {Geanious}}, \bibinfo {author}
  {\bibfnamefont {C.}~\bibnamefont {Ngai}}, \bibinfo {author} {\bibfnamefont
  {C.}~\bibnamefont {Smith}}, \bibinfo {author} {\bibfnamefont
  {S.}~\bibnamefont {Wise}}, \bibinfo {author} {\bibfnamefont {M.~E.}\
  \bibnamefont {Pilgrim}}, \emph {et~al.},\ }\bibfield  {title} {\bibinfo
  {title} {Designing for institutional transformation: Six principles for
  department-level interventions},\ }\href@noop {} {\bibfield  {journal}
  {\bibinfo  {journal} {Physical Review Physics Education Research}\ }\textbf
  {\bibinfo {volume} {15}},\ \bibinfo {pages} {010141} (\bibinfo {year}
  {2019})}\BibitemShut {NoStop}%
\bibitem [{\citenamefont {Dancy}\ \emph {et~al.}(2019)\citenamefont {Dancy},
  \citenamefont {Lau}, \citenamefont {Rundquist},\ and\ \citenamefont
  {Henderson}}]{dancy2019faculty}%
  \BibitemOpen
  \bibfield  {author} {\bibinfo {author} {\bibfnamefont {M.}~\bibnamefont
  {Dancy}}, \bibinfo {author} {\bibfnamefont {A.~C.}\ \bibnamefont {Lau}},
  \bibinfo {author} {\bibfnamefont {A.}~\bibnamefont {Rundquist}},\ and\
  \bibinfo {author} {\bibfnamefont {C.}~\bibnamefont {Henderson}},\ }\bibfield
  {title} {\bibinfo {title} {Faculty online learning communities: A model for
  sustained teaching transformation},\ }\href@noop {} {\bibfield  {journal}
  {\bibinfo  {journal} {Physical Review Physics Education Research}\ }\textbf
  {\bibinfo {volume} {15}},\ \bibinfo {pages} {020147} (\bibinfo {year}
  {2019})}\BibitemShut {NoStop}%
\bibitem [{\citenamefont {Henderson}(2008)}]{henderson2008promoting}%
  \BibitemOpen
  \bibfield  {author} {\bibinfo {author} {\bibfnamefont {C.}~\bibnamefont
  {Henderson}},\ }\bibfield  {title} {\bibinfo {title} {Promoting instructional
  change in new faculty: An evaluation of the physics and astronomy new faculty
  workshop},\ }\href@noop {} {\bibfield  {journal} {\bibinfo  {journal}
  {American Journal of Physics}\ }\textbf {\bibinfo {volume} {76}},\ \bibinfo
  {pages} {179} (\bibinfo {year} {2008})}\BibitemShut {NoStop}%
\bibitem [{\citenamefont {Reinholz}\ \emph {et~al.}(2019)\citenamefont
  {Reinholz}, \citenamefont {Pilgrim}, \citenamefont {Corbo},\ and\
  \citenamefont {Finkelstein}}]{reinholz2019transforming}%
  \BibitemOpen
  \bibfield  {author} {\bibinfo {author} {\bibfnamefont {D.~L.}\ \bibnamefont
  {Reinholz}}, \bibinfo {author} {\bibfnamefont {M.~E.}\ \bibnamefont
  {Pilgrim}}, \bibinfo {author} {\bibfnamefont {J.~C.}\ \bibnamefont {Corbo}},\
  and\ \bibinfo {author} {\bibfnamefont {N.}~\bibnamefont {Finkelstein}},\
  }\bibfield  {title} {\bibinfo {title} {Transforming undergraduate education
  from the middle out with departmental action teams},\ }\href@noop {}
  {\bibfield  {journal} {\bibinfo  {journal} {Change: The Magazine of Higher
  Learning}\ }\textbf {\bibinfo {volume} {51}},\ \bibinfo {pages} {64}
  (\bibinfo {year} {2019})}\BibitemShut {NoStop}%
\bibitem [{\citenamefont {Corbo}\ \emph {et~al.}(2016)\citenamefont {Corbo},
  \citenamefont {Reinholz}, \citenamefont {Dancy}, \citenamefont {Deetz},\ and\
  \citenamefont {Finkelstein}}]{corbo2016framework}%
  \BibitemOpen
  \bibfield  {author} {\bibinfo {author} {\bibfnamefont {J.~C.}\ \bibnamefont
  {Corbo}}, \bibinfo {author} {\bibfnamefont {D.~L.}\ \bibnamefont {Reinholz}},
  \bibinfo {author} {\bibfnamefont {M.~H.}\ \bibnamefont {Dancy}}, \bibinfo
  {author} {\bibfnamefont {S.}~\bibnamefont {Deetz}},\ and\ \bibinfo {author}
  {\bibfnamefont {N.}~\bibnamefont {Finkelstein}},\ }\bibfield  {title}
  {\bibinfo {title} {Framework for transforming departmental culture to support
  educational innovation},\ }\href@noop {} {\bibfield  {journal} {\bibinfo
  {journal} {Physical Review Physics Education Research}\ }\textbf {\bibinfo
  {volume} {12}},\ \bibinfo {pages} {010113} (\bibinfo {year}
  {2016})}\BibitemShut {NoStop}%
\bibitem [{\citenamefont {Lau}\ \emph {et~al.}(2021)\citenamefont {Lau},
  \citenamefont {Martin}, \citenamefont {Corrales}, \citenamefont {Turpen},
  \citenamefont {Goldberg},\ and\ \citenamefont {Price}}]{lau2021taxonomy}%
  \BibitemOpen
  \bibfield  {author} {\bibinfo {author} {\bibfnamefont {A.~C.}\ \bibnamefont
  {Lau}}, \bibinfo {author} {\bibfnamefont {M.}~\bibnamefont {Martin}},
  \bibinfo {author} {\bibfnamefont {A.}~\bibnamefont {Corrales}}, \bibinfo
  {author} {\bibfnamefont {C.}~\bibnamefont {Turpen}}, \bibinfo {author}
  {\bibfnamefont {F.}~\bibnamefont {Goldberg}},\ and\ \bibinfo {author}
  {\bibfnamefont {E.}~\bibnamefont {Price}},\ }\bibfield  {title} {\bibinfo
  {title} {The taxonomy of opportunities to learn (txotl): a tool for
  understanding the learning potential and substance of interactions in faculty
  (online) learning community meetings},\ }\href@noop {} {\bibfield  {journal}
  {\bibinfo  {journal} {International Journal of STEM Education}\ }\textbf
  {\bibinfo {volume} {8}},\ \bibinfo {pages} {1} (\bibinfo {year}
  {2021})}\BibitemShut {NoStop}%
\bibitem [{\citenamefont {McKagan}\ \emph {et~al.}(2020)\citenamefont
  {McKagan}, \citenamefont {Strubbe}, \citenamefont {Barbato}, \citenamefont
  {Mason}, \citenamefont {Madsen},\ and\ \citenamefont
  {Sayre}}]{mckagan2020physport}%
  \BibitemOpen
  \bibfield  {author} {\bibinfo {author} {\bibfnamefont {S.~B.}\ \bibnamefont
  {McKagan}}, \bibinfo {author} {\bibfnamefont {L.~E.}\ \bibnamefont
  {Strubbe}}, \bibinfo {author} {\bibfnamefont {L.~J.}\ \bibnamefont
  {Barbato}}, \bibinfo {author} {\bibfnamefont {B.~A.}\ \bibnamefont {Mason}},
  \bibinfo {author} {\bibfnamefont {A.~M.}\ \bibnamefont {Madsen}},\ and\
  \bibinfo {author} {\bibfnamefont {E.~C.}\ \bibnamefont {Sayre}},\ }\bibfield
  {title} {\bibinfo {title} {Physport use and growth: Supporting physics
  teaching with research-based resources since 2011},\ }\href@noop {}
  {\bibfield  {journal} {\bibinfo  {journal} {The Physics Teacher}\ }\textbf
  {\bibinfo {volume} {58}},\ \bibinfo {pages} {465} (\bibinfo {year}
  {2020})}\BibitemShut {NoStop}%
\bibitem [{\citenamefont {Krane}(2010)}]{krane2010national}%
  \BibitemOpen
  \bibfield  {author} {\bibinfo {author} {\bibfnamefont {K.~S.}\ \bibnamefont
  {Krane}},\ }\bibfield  {title} {\bibinfo {title} {A national workshop in the
  united states to prepare new faculty in physics and astronomy},\ }\href@noop
  {} {\bibfield  {journal} {\bibinfo  {journal} {American Institute of Physics
  Conference Proceedings}\ }\textbf {\bibinfo {volume} {1263}},\ \bibinfo
  {pages} {12} (\bibinfo {year} {2010})}\BibitemShut {NoStop}%
\bibitem [{\citenamefont {Chasteen}\ and\ \citenamefont
  {Chattergoon}(2020)}]{chasteen2020insights}%
  \BibitemOpen
  \bibfield  {author} {\bibinfo {author} {\bibfnamefont {S.~V.}\ \bibnamefont
  {Chasteen}}\ and\ \bibinfo {author} {\bibfnamefont {R.}~\bibnamefont
  {Chattergoon}},\ }\bibfield  {title} {\bibinfo {title} {Insights from the
  physics and astronomy new faculty workshop: How do new physics faculty
  teach?},\ }\href@noop {} {\bibfield  {journal} {\bibinfo  {journal} {Physical
  Review Physics Education Research}\ }\textbf {\bibinfo {volume} {16}},\
  \bibinfo {pages} {020164} (\bibinfo {year} {2020})}\BibitemShut {NoStop}%
\bibitem [{\citenamefont {Bandura}(2012)}]{bandura_functional_2012}%
  \BibitemOpen
  \bibfield  {author} {\bibinfo {author} {\bibfnamefont {A.}~\bibnamefont
  {Bandura}},\ }\href@noop {} {\bibinfo {title} {On the functional properties
  of perceived self-efficacy revisited}} (\bibinfo {year} {2012})\BibitemShut
  {NoStop}%
\bibitem [{\citenamefont {Etel{\"a}pelto}\ \emph {et~al.}(2013)\citenamefont
  {Etel{\"a}pelto}, \citenamefont {V{\"a}h{\"a}santanen}, \citenamefont
  {H{\"o}kk{\"a}},\ and\ \citenamefont {Paloniemi}}]{etelapelto2013agency}%
  \BibitemOpen
  \bibfield  {author} {\bibinfo {author} {\bibfnamefont {A.}~\bibnamefont
  {Etel{\"a}pelto}}, \bibinfo {author} {\bibfnamefont {K.}~\bibnamefont
  {V{\"a}h{\"a}santanen}}, \bibinfo {author} {\bibfnamefont {P.}~\bibnamefont
  {H{\"o}kk{\"a}}},\ and\ \bibinfo {author} {\bibfnamefont {S.}~\bibnamefont
  {Paloniemi}},\ }\bibfield  {title} {\bibinfo {title} {What is agency?
  conceptualizing professional agency at work},\ }\href@noop {} {\bibfield
  {journal} {\bibinfo  {journal} {Educational research review}\ }\textbf
  {\bibinfo {volume} {10}},\ \bibinfo {pages} {45} (\bibinfo {year}
  {2013})}\BibitemShut {NoStop}%
\bibitem [{\citenamefont {Hinostroza}(2020)}]{hinostroza2020university}%
  \BibitemOpen
  \bibfield  {author} {\bibinfo {author} {\bibfnamefont {Y.}~\bibnamefont
  {Hinostroza}},\ }\bibfield  {title} {\bibinfo {title} {University teacher
  educators’ professional agency: A literature review},\ }\href@noop {}
  {\bibfield  {journal} {\bibinfo  {journal} {Professions and Professionalism}\
  }\textbf {\bibinfo {volume} {10}} (\bibinfo {year} {2020})}\BibitemShut
  {NoStop}%
\bibitem [{\citenamefont {Bandura}(2001)}]{bandura_social_2001}%
  \BibitemOpen
  \bibfield  {author} {\bibinfo {author} {\bibfnamefont {A.}~\bibnamefont
  {Bandura}},\ }\bibfield  {title} {\bibinfo {title} {Social cognitive theory:
  An agentic perspective},\ }\href@noop {} {\bibfield  {journal} {\bibinfo
  {journal} {Annual review of psychology}\ }\textbf {\bibinfo {volume} {52}},\
  \bibinfo {pages} {1} (\bibinfo {year} {2001})}\BibitemShut {NoStop}%
\bibitem [{\citenamefont {Thomas}(2008)}]{thomas2008preparing}%
  \BibitemOpen
  \bibfield  {author} {\bibinfo {author} {\bibfnamefont {G.}~\bibnamefont
  {Thomas}},\ }\bibfield  {title} {\bibinfo {title} {Preparing facilitators for
  experiential education: The role of intentionality and intuition},\
  }\href@noop {} {\bibfield  {journal} {\bibinfo  {journal} {Journal of
  Adventure Education \& Outdoor Learning}\ }\textbf {\bibinfo {volume} {8}},\
  \bibinfo {pages} {3} (\bibinfo {year} {2008})}\BibitemShut {NoStop}%
\bibitem [{\citenamefont {English}\ and\ \citenamefont
  {Kitsantas}(2013)}]{english2013supporting}%
  \BibitemOpen
  \bibfield  {author} {\bibinfo {author} {\bibfnamefont {M.~C.}\ \bibnamefont
  {English}}\ and\ \bibinfo {author} {\bibfnamefont {A.}~\bibnamefont
  {Kitsantas}},\ }\bibfield  {title} {\bibinfo {title} {Supporting student
  self-regulated learning in problem-and project-based learning},\ }\href@noop
  {} {\bibfield  {journal} {\bibinfo  {journal} {Interdisciplinary journal of
  problem-based learning}\ }\textbf {\bibinfo {volume} {7}},\ \bibinfo {pages}
  {6} (\bibinfo {year} {2013})}\BibitemShut {NoStop}%
\bibitem [{\citenamefont {Ryan}\ and\ \citenamefont
  {Deci}(2000)}]{ryan2000self}%
  \BibitemOpen
  \bibfield  {author} {\bibinfo {author} {\bibfnamefont {R.~M.}\ \bibnamefont
  {Ryan}}\ and\ \bibinfo {author} {\bibfnamefont {E.~L.}\ \bibnamefont
  {Deci}},\ }\bibfield  {title} {\bibinfo {title} {Self-determination theory
  and the facilitation of intrinsic motivation, social development, and
  well-being.},\ }\href@noop {} {\bibfield  {journal} {\bibinfo  {journal}
  {American psychologist}\ }\textbf {\bibinfo {volume} {55}},\ \bibinfo {pages}
  {68} (\bibinfo {year} {2000})}\BibitemShut {NoStop}%
\bibitem [{\citenamefont {Stupnisky}\ \emph {et~al.}(2018)\citenamefont
  {Stupnisky}, \citenamefont {BrckaLorenz}, \citenamefont {Yuhas},\ and\
  \citenamefont {Guay}}]{stupnisky2018faculty}%
  \BibitemOpen
  \bibfield  {author} {\bibinfo {author} {\bibfnamefont {R.~H.}\ \bibnamefont
  {Stupnisky}}, \bibinfo {author} {\bibfnamefont {A.}~\bibnamefont
  {BrckaLorenz}}, \bibinfo {author} {\bibfnamefont {B.}~\bibnamefont {Yuhas}},\
  and\ \bibinfo {author} {\bibfnamefont {F.}~\bibnamefont {Guay}},\ }\bibfield
  {title} {\bibinfo {title} {Faculty members’ motivation for teaching and
  best practices: Testing a model based on self-determination theory across
  institution types},\ }\href@noop {} {\bibfield  {journal} {\bibinfo
  {journal} {Contemporary Educational Psychology}\ }\textbf {\bibinfo {volume}
  {53}},\ \bibinfo {pages} {15} (\bibinfo {year} {2018})}\BibitemShut {NoStop}%
\bibitem [{\citenamefont {Bandura}(1978)}]{bandura_self-efficacy_1982}%
  \BibitemOpen
  \bibfield  {author} {\bibinfo {author} {\bibfnamefont {A.}~\bibnamefont
  {Bandura}},\ }\bibfield  {title} {\bibinfo {title} {Self-efficacy: Toward a
  unifying theory of behavioral change},\ }\href@noop {} {\bibfield  {journal}
  {\bibinfo  {journal} {Advances in Behaviour Research and Therapy}\ }\textbf
  {\bibinfo {volume} {1}},\ \bibinfo {pages} {139} (\bibinfo {year}
  {1978})}\BibitemShut {NoStop}%
\bibitem [{\citenamefont {Harvey-Jordan}\ and\ \citenamefont
  {Long}(2001)}]{harvey2001process}%
  \BibitemOpen
  \bibfield  {author} {\bibinfo {author} {\bibfnamefont {S.}~\bibnamefont
  {Harvey-Jordan}}\ and\ \bibinfo {author} {\bibfnamefont {S.}~\bibnamefont
  {Long}},\ }\bibfield  {title} {\bibinfo {title} {The process and the pitfalls
  of semi-structured interviews},\ }\href@noop {} {\bibfield  {journal}
  {\bibinfo  {journal} {Community Practitioner}\ }\textbf {\bibinfo {volume}
  {74}},\ \bibinfo {pages} {219} (\bibinfo {year} {2001})}\BibitemShut
  {NoStop}%
\bibitem [{\citenamefont {Yin}(2009)}]{yin2009case}%
  \BibitemOpen
  \bibfield  {author} {\bibinfo {author} {\bibfnamefont {R.~K.}\ \bibnamefont
  {Yin}},\ }\bibfield  {title} {\bibinfo {title} {Case study research: Design
  and methods},\ }\href@noop {} {\bibfield  {journal} {\bibinfo  {journal}
  {sage}\ }\textbf {\bibinfo {volume} {5}} (\bibinfo {year}
  {2009})}\BibitemShut {NoStop}%
\bibitem [{\citenamefont {Creswell}\ and\ \citenamefont
  {Poth}(2016)}]{creswell2016qualitative}%
  \BibitemOpen
  \bibfield  {author} {\bibinfo {author} {\bibfnamefont {J.~W.}\ \bibnamefont
  {Creswell}}\ and\ \bibinfo {author} {\bibfnamefont {C.~N.}\ \bibnamefont
  {Poth}},\ }\bibfield  {title} {\bibinfo {title} {Qualitative inquiry and
  research design: Choosing among five approaches},\ }\href@noop {} {\bibfield
  {journal} {\bibinfo  {journal} {Sage publications}\ } (\bibinfo {year}
  {2016})}\BibitemShut {NoStop}%
\bibitem [{\citenamefont {Suri}(2011)}]{suri2011purposeful}%
  \BibitemOpen
  \bibfield  {author} {\bibinfo {author} {\bibfnamefont {H.}~\bibnamefont
  {Suri}},\ }\bibfield  {title} {\bibinfo {title} {Purposeful sampling in
  qualitative research synthesis},\ }\href@noop {} {\bibfield  {journal}
  {\bibinfo  {journal} {Qualitative research journal}\ } (\bibinfo {year}
  {2011})}\BibitemShut {NoStop}%
\bibitem [{\citenamefont {Emmel}(2013)}]{emmel2013purposeful}%
  \BibitemOpen
  \bibfield  {author} {\bibinfo {author} {\bibfnamefont {N.}~\bibnamefont
  {Emmel}},\ }\bibfield  {title} {\bibinfo {title} {Sampling and choosing cases
  in qualitative research: A realist approach},\ }\href@noop {} {\bibfield
  {journal} {\bibinfo  {journal} {Sage}\ } (\bibinfo {year}
  {2013})}\BibitemShut {NoStop}%
\bibitem [{\citenamefont {Nissen}\ and\ \citenamefont
  {Shemwell}(2016)}]{nissen2016gender}%
  \BibitemOpen
  \bibfield  {author} {\bibinfo {author} {\bibfnamefont {J.~M.}\ \bibnamefont
  {Nissen}}\ and\ \bibinfo {author} {\bibfnamefont {J.~T.}\ \bibnamefont
  {Shemwell}},\ }\bibfield  {title} {\bibinfo {title} {Gender, experience, and
  self-efficacy in introductory physics},\ }\href@noop {} {\bibfield  {journal}
  {\bibinfo  {journal} {Physical Review Physics Education Research}\ }\textbf
  {\bibinfo {volume} {12}},\ \bibinfo {pages} {020105} (\bibinfo {year}
  {2016})}\BibitemShut {NoStop}%
\bibitem [{\citenamefont {Kalender}\ \emph {et~al.}(2020)\citenamefont
  {Kalender}, \citenamefont {Marshman}, \citenamefont {Schunn}, \citenamefont
  {Nokes-Malach},\ and\ \citenamefont {Singh}}]{kalender2020damage}%
  \BibitemOpen
  \bibfield  {author} {\bibinfo {author} {\bibfnamefont {Z.~Y.}\ \bibnamefont
  {Kalender}}, \bibinfo {author} {\bibfnamefont {E.}~\bibnamefont {Marshman}},
  \bibinfo {author} {\bibfnamefont {C.~D.}\ \bibnamefont {Schunn}}, \bibinfo
  {author} {\bibfnamefont {T.~J.}\ \bibnamefont {Nokes-Malach}},\ and\ \bibinfo
  {author} {\bibfnamefont {C.}~\bibnamefont {Singh}},\ }\bibfield  {title}
  {\bibinfo {title} {Damage caused by women’s lower self-efficacy on physics
  learning},\ }\href@noop {} {\bibfield  {journal} {\bibinfo  {journal}
  {Physical Review Physics Education Research}\ }\textbf {\bibinfo {volume}
  {16}},\ \bibinfo {pages} {010118} (\bibinfo {year} {2020})}\BibitemShut
  {NoStop}%
\bibitem [{\citenamefont {Hass}\ and\ \citenamefont
  {El-Adawy}(2022)}]{hass2022emerging}%
  \BibitemOpen
  \bibfield  {author} {\bibinfo {author} {\bibfnamefont {C.}~\bibnamefont
  {Hass}}\ and\ \bibinfo {author} {\bibfnamefont {S.}~\bibnamefont
  {El-Adawy}},\ }\bibfield  {title} {\bibinfo {title} {Emerging mathematics
  education researchers' conception of theory in education research},\
  }\href@noop {} {\bibfield  {journal} {\bibinfo  {journal} {Proceedings of the
  Annual Conference on Research in Undergraduate Mathematics Education}\ }
  (\bibinfo {year} {2022})}\BibitemShut {NoStop}%
\end{thebibliography}%

\end{document}